\documentclass[a4paper]{scrartcl}

\usepackage{lmodern}
\usepackage{amssymb,amsmath}
\usepackage[T1]{fontenc}
\usepackage[utf8]{inputenc}
\usepackage{upquote}
\usepackage[tracking=smallcaps,letterspace=70]{microtype}
\usepackage[unicode=true]{hyperref}
\usepackage{longtable,booktabs}
\usepackage{graphicx,grffile}
\usepackage{authblk}
\usepackage[comma,super,sort&compress]{natbib}
\usepackage{tabularx}
\newcolumntype{Y}{>{\centering\arraybackslash}X}
\usepackage[font=small]{caption}
\usepackage{multirow}

\usepackage[british]{babel}
\usepackage[modulo]{lineno}

\UseMicrotypeSet[protrusion]{basicmath}

\providecommand{\subtitle}[1]{}

\PassOptionsToPackage{usenames,dvipsnames}{color}

\hypersetup{%
	pdftitle={On flow disturbances caused by pressure taps in highly elastic flows around a microfluidic cylinder},
	pdfauthor={R. Rodrigues; T. Rodrigues; L. Campo-Dea{\~n}o},
	pdfkeywords={Keyword \#1, keyword \#2, keyword \#3},
	pdfborder={0 0 0},
	colorlinks=true,
	linkcolor=Blue,
	citecolor=Blue,
	urlcolor=Blue,
	breaklinks=true
}%

\makeatletter
	\def\maxwidth{\ifdim\Gin@nat@width>\linewidth\linewidth\else\Gin@nat@width\fi}
	\def\maxheight{\ifdim\Gin@nat@height>\textheight\textheight\else\Gin@nat@height\fi}
\makeatother

\setkeys{Gin}{width=\maxwidth,height=\maxheight,keepaspectratio}

\ifx\paragraph\undefined\else
	\let\oldparagraph\paragraph
	\renewcommand{\paragraph}[1]{\oldparagraph{#1}\mbox{}}
\fi
\ifx\subparagraph\undefined\else
	\let\oldsubparagraph\subparagraph
	\renewcommand{\subparagraph}[1]{\oldsubparagraph{#1}\mbox{}}
\fi

\title{On flow disturbances caused by pressure taps in highly elastic flows around a microfluidic cylinder}

\author[ ]{R. Rodrigues}
\author[ ]{T. Rodrigues}
\author[$\dagger$]{L. Campo-Dea{\~n}o}

\affil[ ]{CEFT - Centro de Estudos de Fen{\'o}menos de Transporte, Depto.~de Engenharia Mec{\^a}nica, Faculdade de Engenharia, Universidade do Porto, Rua Dr.\ Roberto Frias, 4200-465, Porto, Portugal} 
\affil[ ]{ALiCE - Laborat{\'o}rio Associado em Engenharia Qu{\'i}mica, Faculdade de Engenharia, Universidade do Porto, Rua Dr.\ Roberto Frias, 4200-465, Porto, Portugal}
\affil[$\dagger$]{Email: \href{mailto:campo@fe.up.pt}{campo@fe.up.pt}}

\date{}

\begin{document}

\maketitle

\begin{abstract} 
\small
	\noindent The objective of this work is to characterise the onset of laterally asymmetric flow of viscoelastic solutions around a confined microfluidic cylinder, which was encountered in a recent study [Rodrigues et al., \textit{J. Non-Newton. Fluid Mech.} \textbf{286}, 104406 (2020)]. To this end, two non-Newtonian fluids were employed in the same micro-geometry. Two microchannels were studied, both with a cylinder of diameter 75 \textmu m, aspect ratio (channel height over width) of 0.37 and blockage ratio (cylinder diameter over channel width) of 0.28, differing only on the width of the pressure taps, located 500 \textmu m up- and downstream from the respective cylinder face, on opposing walls. The working fluids consist of two poly(ethylene oxide) (PEO) solutions: a weakly shear-thinning elastic fluid and an elastic shear-thinning fluid. Micro-Particle Image Velocimetry (\textmu PIV) and streak imaging techniques were used to evaluate the flow over a Weissenberg number range: \mbox{$100\leq Wi\leq500$}, while maintaining a low Reynolds number, $Re<1$. The elastic shear-thinning solution showed laterally asymmetric flow past the cylinder with both pressure tap designs, while with the weakly shear-thinning solution asymmetric flow was only observed with the wider pressure tap intake. In both cases, the fluids preferentially chose the cylinder/wall gap opposing the upstream pressure tap, which was found to influence the flow greatly, seemingly associated with time-dependent flow and possibly the lateral flow asymmetry itself. This work brings to light the necessary compromise between optimal pressure tap design for quality pressure measurements and minimal flow interference, due to the increased susceptibility of elastic microfluidic flows to flow perturbations.\vspace{.5cm}
 
\end{abstract}

\newpage

\section{Introduction}
\label{sec:intro}

A few-micron, untethered microbot deployed into the blood stream could have access to the entirety of the human body, allowing for medical imaging, drug delivery, diagnosis, and surgery to be conducted locally, avoiding more traumatic and invasive procedures. However, the blood flow around swimming microbots is still to be thoroughly understood \cite{martinez2016}. This subject was recently tackled by Rodrigues et al.\citep{rodrigues2020} who simulated the haemodynamics around a swimming microbot through the benchmark \textit{flow past a confined cylinder}, described in-depth in their work. The authors set out to study the steady flow (from creeping flow to moderate Reynolds numbers) of polymeric blood analogues (one Newtonian and five weakly elastic shear-thinning fluids) around two different microfluidic cylinders, one circular and one elliptic, confined in a small-blood-vessel-like microchannel. The microchannels employed by the authors had pressure taps incorporated \mbox{500 \textmu m} up- and downstream of the cylinder, which were designed, and tested with a Newtonian fluid (de-ionised water) yielding reliable pressure measurements, in their previous work\citep{rodrigues2019}. Focusing on the results obtained with the circular cylinder, Rodrigues et al.\citep{rodrigues2020} reported streamline divergence upstream of the cylinder and stronger streamline curvature downstream with the viscoelastic solutions compared to the Newtonian blood analogue. Moreover, extended downstream wakes were also found, and were aggravated with fluid elasticity. An absence of downstream flow separation and vorticity was also noted for all viscoelastic fluids. With their most elastic fluid, a slight lateral flow asymmetry was found for a viscoelastic Mach number (representing the ratio of the local flow speed to the velocity of the viscoelastic waves) $Ma\approx1$, and it decayed with increasing flow rate as inertial effects started to dominate. The authors mention the similarity of this flow asymmetry with a previously reported elastic instability\citep{haward2020}. 

Viscoelastic instabilities are no novelty in non-Newtonian microfluidic flows and, in the past few years, there have been significant efforts in the study of viscoelastic flows around microfluidic cylinders, focusing on elastic instabilities, which provided several interesting observations. Kenney et al.\citep{kenney2013}, Galindo-Rosales et al.\citep{galindo2014} and Shi et al.\citep{shi2015} conducted experimental work on Boger fluids, encountering an instability upstream of the cylinder, near the forward stagnation point, visible through a ``pinch'' in the streamlines, resembling a wineglass shape. This instability was found, in each case, for relatively high blockage ratios (cylinder diameter over channel width: $\beta\geq0.5$). Kenney et al.\citep{kenney2013} found that increasing the Reynolds number led to an aggravation of the ``pinch'', incorporating more streamlines, reducing the width of the wineglass-shaped flow and disconnecting the forward stagnation point from the cylinder face further upstream. A vortex formed between the detached stagnation point and the cylinder face, which remained steady until further increasing the flow rate, leading to a cycle of vortex growth and collapse, giving the forward stagnation point a time-dependent axial movement. It was generally found that increasing confinement (decreasing the aspect ratio and increasing the channel blockage) and fluid elasticity led to an anticipation of the onset of this instability. Moreover, Kenney et al.\citep{kenney2013} concluded that the instability was of purely elastic origin, but inertia still played a role as increasing the Reynolds number enhanced the behaviour, and less elastic solutions were still able to manifest the instability at sufficiently high Reynolds numbers. Kenney et al.\citep{kenney2013} ascribe its onset primarily to enhanced extensional stresses and the large confinement and blockage of the channel. Significantly enlarged downstream wakes were also reported, a common characteristic of viscoelastic flows together with vortex shedding suppression. Haward et al.\cite{haward2018} also studied the flow of a practically non-shear-thinning polymeric solution around confined cylinders and noted extremely long downstream wakes despite not reporting the wineglass instability. The authors attribute the length of the wake to the deformation of the polymer molecules in a strong extensional flow region upstream of the cylinder that then advect around the cylinder face and flow into the downstream wake where they begin to relax. Additionally, they noted a very slight lateral flow asymmetry upstream of the cylinder. 

In the last few years, other instabilities upstream of the cylinder have been reported with shear-thinning fluids. Nolan et al.\cite{nolan2016} analysed the flow of weakly elastic, shear-thinning solutions, encountering extended downstream wakes and detachment of the forward stagnation point into further upstream, similar to the findings of Kenney et al.\citep{kenney2013} with Boger fluids. Additionally, a buckling-type instability was found immediately upstream of the cylinder, ascribed to viscosity fluctuations near the forward stagnation point where a slow-moving high-viscosity stream collided with the faster-moving surrounding fluid. The instability was reported to be enhanced by fluid elasticity and a viscoelastic Mach number, $Ma=1$, was found to be representative of its onset, while, as the Reynolds number increased, inertia led to the instability's decay. Decoupling fluid elasticity and shear-thinning, Ribeiro et al.\cite{ribeiro2014} tested the flow of three non-Newtonian polymeric solutions, one shear-thinning and two Boger fluids, past a cylinder with \mbox{10 mm} of diameter (with a blockage ratio of 0.5). Despite being conducted at the macroscale, diverging from the remaining works discussed here, we still consider the observations of Ribeiro et al.\cite{ribeiro2014} to be noteworthy. They encountered the lateral flow asymmetry with the shear-thinning fluid through abnormal streamline bending at the forward stagnation point of their cylinder, whereas the constant-viscosity solutions showed no evidence of the flow asymmetry, nor the streamline ``pinch'' reported in other Boger fluid studies. They additionally tested several channel geometries, finding that decreasing the channel aspect ratio delayed the onset of the instability due to the viscous effects at the channel walls.

Zhao et al.\cite{zhao2016} and Hopkins et al. \citep{hopkins2022} both conducted experiments with strongly shear-thinning wormlike micellar solutions (WLMs) testing the effects of the channel blockage. Zhao et al.\cite{zhao2016} tested several blockage ratios (from 0.5 to 0.83) and found a sequence of flow transitions while increasing the flow rate. The Newtonian-like flow was compromised firstly by bending streamlines upstream of the cylinder, agreeing with the observations of Ribeiro et al.\cite{ribeiro2014}, and further increasing the flow rate led to the detachment of the forward stagnation point and the growth of a vortex immediately before the cylinder edge, similar to the findings of Kenney et al.\cite{kenney2013} but with no streamline ``pinch''. Thereafter, unsteady flow onset, first downstream and then on the whole flow. The authors also found that increasing the blockage ratio facilitated the onset of these phenomena. Recently, Hopkins et al. \citep{hopkins2022} studied a wide range of blockage ratios (from 0.055 to 0.63) over a large Weissenberg number interval (representing the ratio of elastic to viscous forces, up to 1000), while maintaining low Reynolds numbers. The employed channel aimed to approach 2D flow with a high aspect ratio. The authors found different sets of flow transitions, depending on the degree of channel blockage, but despite the differences between both flows, two critical Weissenberg numbers could be defined. Above the first critical Weissenberg number, a time-steady, lateral flow asymmetry onset, while above the second critical number, time-dependent flow was observed. A low blockage ratio was considered smaller than 0.33, where the associated instabilities are linked to the downstream stagnation point. Above 0.41, it was considered a high blockage ratio and the instabilities were connected to the forward stagnation point instead (in accordance with the results obtained by Zhao et al.\citep{zhao2016}), with bending streamlines and wall-attached vortices upstream of the cylinder (studied in-depth in their previous work \cite{hopkins2022upstream}). In the intermediate blockage range, between 0.33 and 0.41, the flow dynamics are more complex and flow transitions less defined, possibly due to the competition between the instabilities arising from either stagnation point. For either low or high blockage ratios, increasing the channel blockage seemed to decrease both critical Weissenberg numbers (also agreeing with Zhao et al.\citep{zhao2016}).

Haward et al.\cite{haward2019} utilised rigid, high-aspect- and low-blockage-ratio (5 and 0.1, respectively) channels to approximate creeping two-dimensional flow, also with a strongly shear-thinning WLM. The authors first noticed the development of a downstream wake and then, increasing the Weissenberg number, lateral flow asymmetry onset and grew significantly until virtually all of the fluid was flowing through one of the cylinder/wall gaps. Thereafter, time-dependent flow was noticed, and further increasing the Weissenberg number led to a partial resymmetrisation of the flow. The authors mentioned that the preferential path around the cylinder was chosen seemingly at random and introduced an asymmetry parameter to quantify the degree of asymmetry in the flow, $I = \lvert u_i-u_j \rvert/(u_i+u_j)$, where $u_i$ and $u_j$ are the streamwise velocities at each gap centre-point. They also presented an interesting temporal analysis of the asymmetry once time-dependent flow onset. The authors present the possibility of the onset of the asymmetric flow being due to a random fluctuation in the downstream high-elongational-stress wake region, leading to a momentary imbalance of flow rates, and thus shear rates, in the cylinder/wall gaps, which is then maintained and escalated by the strong shear-thinning behaviour of the fluid. The partial resymmetrisation of the flow was initially thought to be a consequence of the breakage of WLM chains until their posterior work proved otherwise\citep{haward2020}. In this study, Haward et al.\cite{haward2020} studied several polymeric shear-thinning fluids  and the results again showed extended downstream wakes and lateral flow asymmetry, once again aggravating with flow rate until regaining partial symmetry, which proved that the phenomenon was not a particular behaviour of WLMs, but a more general effect. The authors mention the possible necessity of both fluid elasticity and shear-thinning for the onset and feeding mechanisms of the instability. 

After the encounter of this lateral flow asymmetry by Haward et al.\citep{haward2019,haward2020}, Varchanis et al.\citep{varchanis2020} set out to compute numerically, and confirm experimentally, 2D simulations of creeping flow of three different fluid types: an inelastic, shear-thinning fluid; a non-shear-thinning, elastic fluid and an elastic, shear-thinning fluid. The microfluidic geometry was the same as was employed in the studies of Haward et al.\citep{haward2018,haward2019, haward2020}. The flow remained essentially symmetric for the first, inelastic, shear-thinning fluid, either laterally or fore-aft. For the second, non-shear-thinning fluid, extended wakes were reported and time-dependent flow was found numerically after a critical Weissenberg number, with lateral oscillations of the downstream wake connected to a small transient lateral asymmetry around the cylinder, but that converged to symmetric flow when time-averaging. This slight, time-dependent lateral flow asymmetry agrees with the results obtained by Haward et al.\cite{haward2018} with a practically non-shear-thinning elastic fluid. There was a good agreement between the simulation and the experimental results for the final, elastic and shear-thinning fluid, showing downstream wakes and lateral asymmetry. The preferential path of choice around the cylinder was chosen by the fluid seemingly at random, as was previously reported by Haward et al. \citep{haward2020,haward2019}. The numerical simulations reported an increasing bending of the downstream wake with increasing flow rate, following the flow asymmetry. The authors concluded numerically that the independent increase in shear-thinning and elasticity resulted similarly in facilitating the onset of the asymmetric flow and increasing the maximum asymmetry achievable. Furthermore, both elasticity and shear-thinning were found to be necessary conditions for the onset, and the increasing blockage of the channel facilitated the asymmetry's onset and maintenance to a degree. Increasing the channel blockage past a certain limit, the flow begins to regain symmetry as the intense wall shear brings the fluid's viscosity near the high-shear Newtonian plateau.

The numerical study of highly elastic flows is a strenuous endeavour, partly due to numerical instabilities, namely the high Weissenberg number problem (HWNP) \citep{varchanis2020,peng2023}, which made, to this day, the simulation of viscoelastic microfluidic flows still a somewhat recent subject. Varchanis et al.\citep{varchanis2020} were able to avoid the HWNP, successfully simulating the lateral-asymmetry instability and, in a more recent paper, Pen et al.\citep{peng2023} present a fascinating numerical study of a viscoelastic instability upstream of circular cylinders at a low Reynolds and relatively high Weissenberg numbers. The development of these numerical methods allows for a further in-depth investigation of elastic flows and may shed some light on the mechanisms of viscoelastic instabilities previously reported in experimental works.

Our goal is to study the onset of the aforementioned lateral flow asymmetry in two microchannels, both consisting of the same general geometry (the same as employed by Rodrigues et al.\citep{rodrigues2020} in their circular cylinder study), differing only on the pressure tap designs. To this end, an elastic weakly shear-thinning, and an elastic shear-thinning fluids (the same employed by Varchanis et al.\citep{varchanis2020}) are studied over five flow regimes, which are defined by Weissenberg numbers from 100 to 500. Micro-particle Image Velocimetry (\textmu PIV) is used to obtain the velocity vector fields, from which local viscosity and flow-type parameter flow fields are computed and analysed. The streamwise velocity profiles and asymmetry parameters, similar to the ones used in the work of Haward et al.\citep{haward2019,haward2020} and Varchanis et al.\citep{varchanis2020}, are also evaluated. Finally, fluorescent streak imaging gives us qualitative information regarding the flow patterns.

\section{Materials and methods}
\label{sec:methods}

\subsection{Test fluids}
\label{sec:methods:test fluids}

The test fluids consisted of two poly(ethylene oxide) (PEO) aqueous solutions. The weakly shear-thinning fluid (PEO4) was composed of \mbox{2000 ppm} of 4 MDa PEO dissolved in a mixture of \mbox{50 wt\%} glycerol and water, and the shear-thinning fluid (PEO8) \mbox{5000 ppm} of 8 MDa PEO dissolved in pure water. Additionally, \mbox{50 ppm} of sodium azide (SA) was added to both solutions, protecting them from microorganisms not influencing the rheological properties of either fluid at this concentration. The densities of both fluids were obtained with a pycnometer at room temperature (approximately 20$^\circ$C): \mbox{$\rho_\mathrm{PEO4}$ = $1125 \pm 1.2$} and \mbox{$\rho_\mathrm{PEO8} = 998 \pm 0.1$ kg/m$^3$}. 

The steady shear viscosity curves were obtained using a stress-controlled rotational rheometer, Anton Paar Physica MCR 301 (minimum torque \mbox{$T_\mathrm{min} = 0.1$ \textmu N$\cdot$m}) equipped with a parallel-plate geometry of radius \mbox{$R = 25$ mm}. The temperature was kept constant at 20$^\circ$C and the gap, $h$, was set to \mbox{100 \textmu m}.

Several phenomena can negatively affect the viscosity measurements; mechanical limitations, secondary flows and viscoelastic instabilities are known causes for inducing error in apparent viscosity\citep{ewoldt2015}. The low torque limit, beneath which the rheometer cannot conduct accurate measurements, is given, as a function of the shear rate, $\Dot{\gamma}$, by\citep{ewoldt2015}:

\begin{equation}
    \eta_{\mathrm{min}}^{\mathrm{torque}} = 
    \frac{2 T_{\mathrm{min}}}{\pi R^3 \Dot{\gamma}}\,.
\end{equation}

Secondary flows can be expected at high shear rates due to fluid inertia and generate an apparent viscosity increase. For the parallel-plate geometry, secondary flows can be expected above the limit given by:

\begin{equation}
    \eta_\mathrm{max}^\mathrm{sec} = \frac{\rho h^3 \Dot{\gamma}}{Re_\mathrm{crit}R}\,,
\end{equation}
where $Re_\mathrm{crit}$ is the critical Reynolds number above which we can expect secondary flows and, for a torque error of 1\%, it takes 4 as an approximate value\citep{ewoldt2015}.

\begin{figure}[htp]
    \centering
    \includegraphics[trim={0cm 0cm 0cm 0cm},clip,width=.9\textwidth]{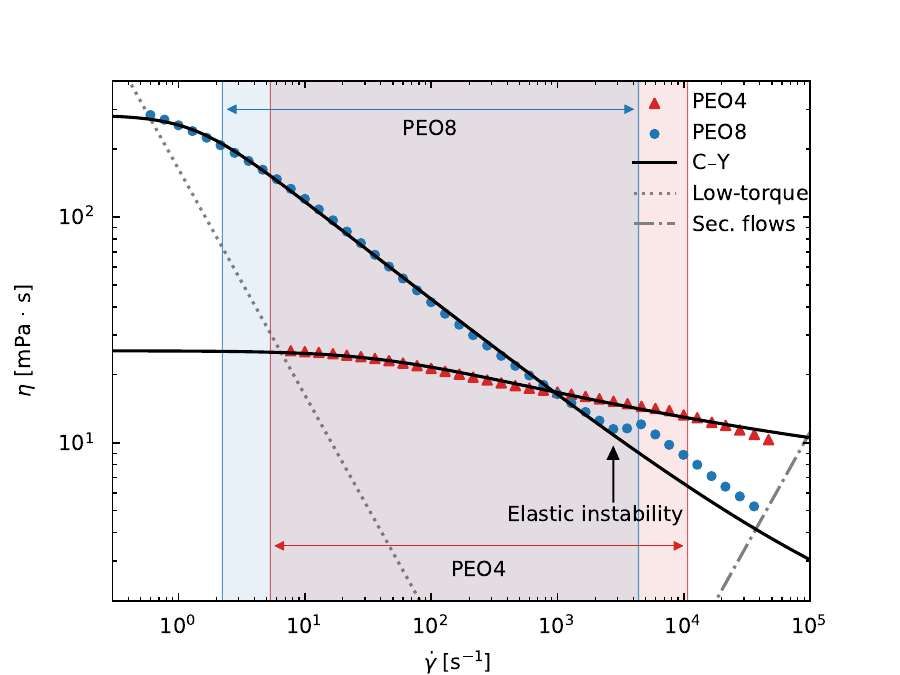}
    \caption{Steady shear viscosity curves (obtained with the parallel-plate geometry, \mbox{$R = 25$ mm}, at 20$^\circ$C), relevant experimental limits, Carreau--Yasuda viscosity models and experimental range of shear rates achieved in our work (vertical bands, plotted from the local shear rates, which were calculated from \textmu PIV data), for each fluid.}
    \label{fig:useful}
\end{figure}

Figure \ref{fig:useful} shows the viscosity curves for both fluids and experimental limits. PEO4 presents a very slight shear-thinning, practically negligible when compared with PEO8, which, in turn, shows a very clear viscosity decrease with shear rate. A slight increase in apparent viscosity was noticed for PEO8 (at \mbox{$\Dot{\gamma} \approx 2770$ s$^{-1}$}), which was attributed to a possible elastic instability. Oliveira et al.\citep{oliveira2006} reported a similar behaviour with a PEO solution, quantifying their instability's onset through the Reynolds and Weissenberg numbers: \mbox{$Re = (\rho \Omega R^2\theta_0)/\eta(\Dot{\gamma}) = 8.3$} and \mbox{$Wi = \lambda\Dot{\gamma}\sqrt{\theta_0} \approx 11$} (where $\theta_0$ is the cone-plate angle). Similarly characterising our instability (having \mbox{$\theta_0 \approx h/R$})\citep{pakdel1996}: $Re=2.40$ and $Wi=23.30$.

Carreau--Yasuda (C--Y) viscosity models were chosen for each fluid, given by:

\begin{equation}
    \eta_{\mathrm{C-Y}} = \eta_\infty + (\eta_0 - \eta_\infty)\left[1+\left(\frac{\Dot{\gamma}}{\Dot{\gamma}^*}\right)^a \right]^\frac{n-1}{a}\,,
    \label{eq:CY}
\end{equation}
where $\eta_\infty$ is the infinite-shear viscosity, $\eta_0$ is the zero-shear viscosity, $\Dot{\gamma}^*$ is the characteristic shear rate for the onset of shear-thinning, $n$ is the dimensionless power-law index and $a$ is a dimensionless exponent that describes the transition from the zero-shear plateau to the power-law region \citep{haward2020}. The fitting process was done by minimising the following error function\citep{escudier2001}:

\begin{equation}
    \sum\limits_{N} \left( 1 - \frac{\eta_{\mathrm{M}}}{\eta_{\mathrm{C-Y}}} \right)^2\,,
\end{equation}
where $\eta_{\mathrm{M}}$ and $\eta_{\mathrm{C-Y}}$ are the measured and the C--Y viscosities, respectively.

\begin{figure}[htp]
    \centering
    \includegraphics[trim={0cm 0cm 0cm 0cm},clip,width=.9\textwidth]{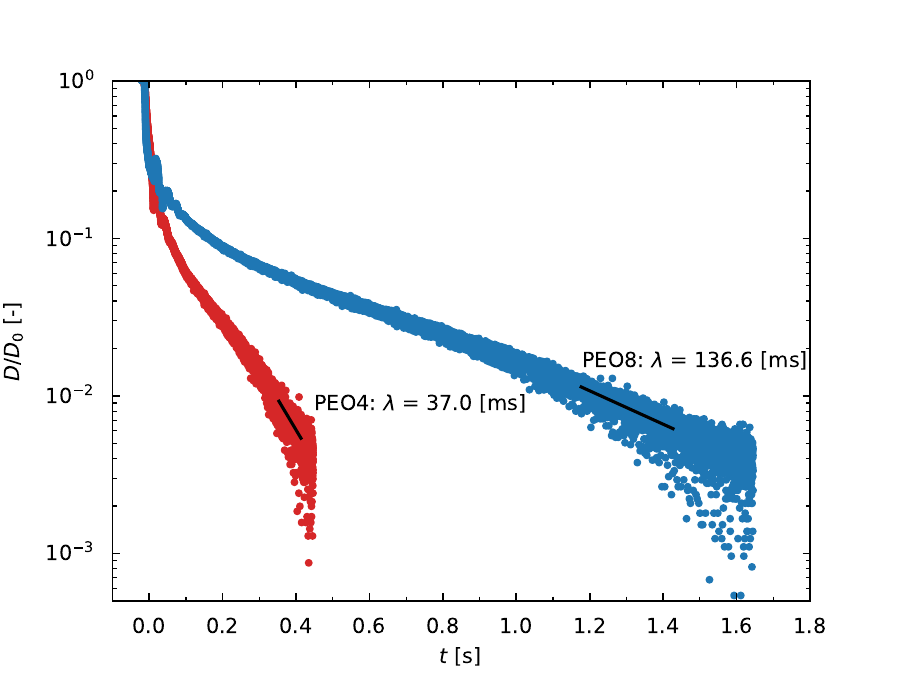}
    \caption{Filament diameter as a function of time for PEO4 and PEO8.}
    \label{fig:caber}
\end{figure}

The extensional relaxation time of a fluid, $\lambda$, can be obtained by quantifying the variation of the diameter of a liquid bridge with time in the elasto-capillary regime. This was done through capillary breakup extensional rheometry with a Thermo Scientific HAAKE$^{\mathrm{TM}}$ CaBER$^{\mathrm{TM}}$ 1. To obtain reliable results, the initial height, $h_0$, must be sufficiently small to sustain the liquid bridge, ensuring that surface tension is capable of supporting the fluid filament against gravitational collapse\citep{rodd2005}: \mbox{$h_0/l_\mathrm{cap}<1$}, where \mbox{$l_\mathrm{cap}$} is the capillary length (dependent on surface tension, fluid density and gravitational acceleration: \mbox{$l_\mathrm{cap} = \sqrt{\sigma/\rho g}$}). Furthermore, numerical studies have found an optimal range for the initial aspect ratio\citep{yao1998}: \mbox{$0.5\leq h_0/D_0\leq1$}, where $D_0$ is the measuring geometry's diameter. We chose initial and final heights of \mbox{$h_0 = 2$} and \mbox{$h_\mathrm{f} = 6.5$ mm}, respectively, and a measuring geometry diameter \mbox{$D_0 = 4$ mm} for both fluids. All tests were conducted at room temperature, with  linear plate separation and a strike time of \mbox{20 ms}. From the time variation of the filaments' diameter, $D(t)$, we can then calculate the relaxation time of the fluid through an exponential fit in the elasto-capillary regime. The fitting function is given by\citep{mckinley2005}: 

\begin{equation}
    \frac{D(t)}{D_0} \approx \left(\frac{D_0 G}{4\sigma} \right)^{1/3} \exp\left(-\frac{t}{3\lambda}\right)\,,
    \label{eq:relax}
\end{equation}
where $G$ is the fluid’s elastic modulus. In Figure \ref{fig:caber} are shown
two of the data sets, one for each fluid, given by CaBER and their respective exponential fits and calculated relaxation times.

For each fluid, over ten measurements were done to ensure replicability. The results yielded relaxation times of: \mbox{$\lambda_\mathrm{PEO4} = 38.3\pm3.4$ ms} and \mbox{$\lambda_\mathrm{PEO8} = 133.0\pm8.2$ ms}.

Table \ref{tab:fluids} shows the measured densities, the obtained relaxation times and Carreau--Yasuda material constants for each fluid.

\begin{table}[htp]
    \centering
    \renewcommand{\arraystretch}{1.5}
    \small{
    \caption{Densities, relaxation times and Carreau--Yasuda material constants of PEO4 and PEO8}
    \label{tab:fluids}
    \begin{tabularx}{1\textwidth}{YYYYYYYY}
    \hline
    \multirow{2}{*}{Solution}& $\rho$&
    $\lambda$&
    $ \eta_{0}$& $ \eta_{\infty}$& $ \Dot{\gamma}^*$ & $ n$&
    $ a$\\
    &[kg/m$ ^3$]&
    [ms]&
    [mPa$ \cdot$s]&
    [mPa$ \cdot$s]&
    [s$ ^{-1}$]&
    [-]&
    [-]\\
    \hline
    PEO4&
    1125&
    38.3& 
    25.6&
    5.9&
    40.8&
    0.8&
    1.1\\
    \hline
    PEO8&
    998&
    133.0&
    282.6&
    1.0&
    1.3&
    0.6&
    1.9\\
    \hline
    \end{tabularx}}
\end{table}

\subsection{Microfluidic devices design and fabrication}
\label{sec:methods:devices}

The microchannels consisted of a \mbox{10 mm} long channel with rectangular cross-section of width \mbox{$2L_\mathrm{c} = 270$ \textmu m}, and depth \mbox{$H = 100$ \textmu m}. A circular cylinder of diameter \mbox{$D = 2R = 75$ \textmu m} was placed at the channel centre, \mbox{7 mm} from the channel inlet, free from entrance effects. The most significant characteristics of the channel are its aspect and blockage ratios, \mbox{$\alpha = H/2L_\mathrm{c} = 0.37$} and \mbox{$\beta = D/2L_\mathrm{c} = 0.28$}, respectively, and the cylinder's aspect ratio, \mbox{$\Lambda = H/D = \alpha/\beta = 1.33$}.

\begin{figure}[htp]
    \centering
    \begin{minipage}{.45\textwidth}
    \centering
    \small
    (a) WTI ($w_\mathrm{WTI}=108$ \textmu m)
    \includegraphics[width=\textwidth]{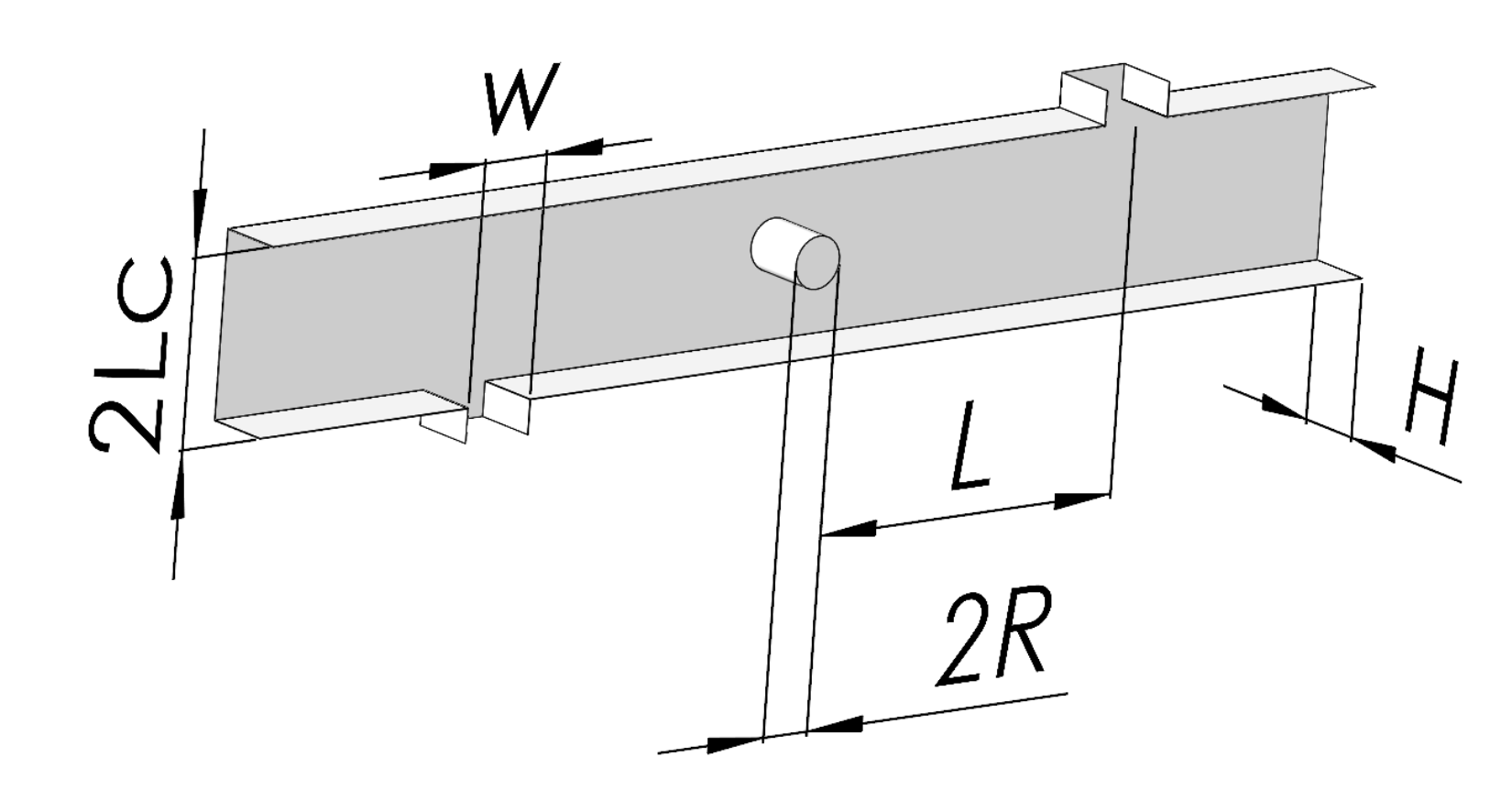}
    \end{minipage}
    \begin{minipage}{.45\textwidth}
    \centering
    \small
    (b) NTI ($w_\mathrm{NTI}=54$ \textmu m)
    \includegraphics[width=\textwidth]{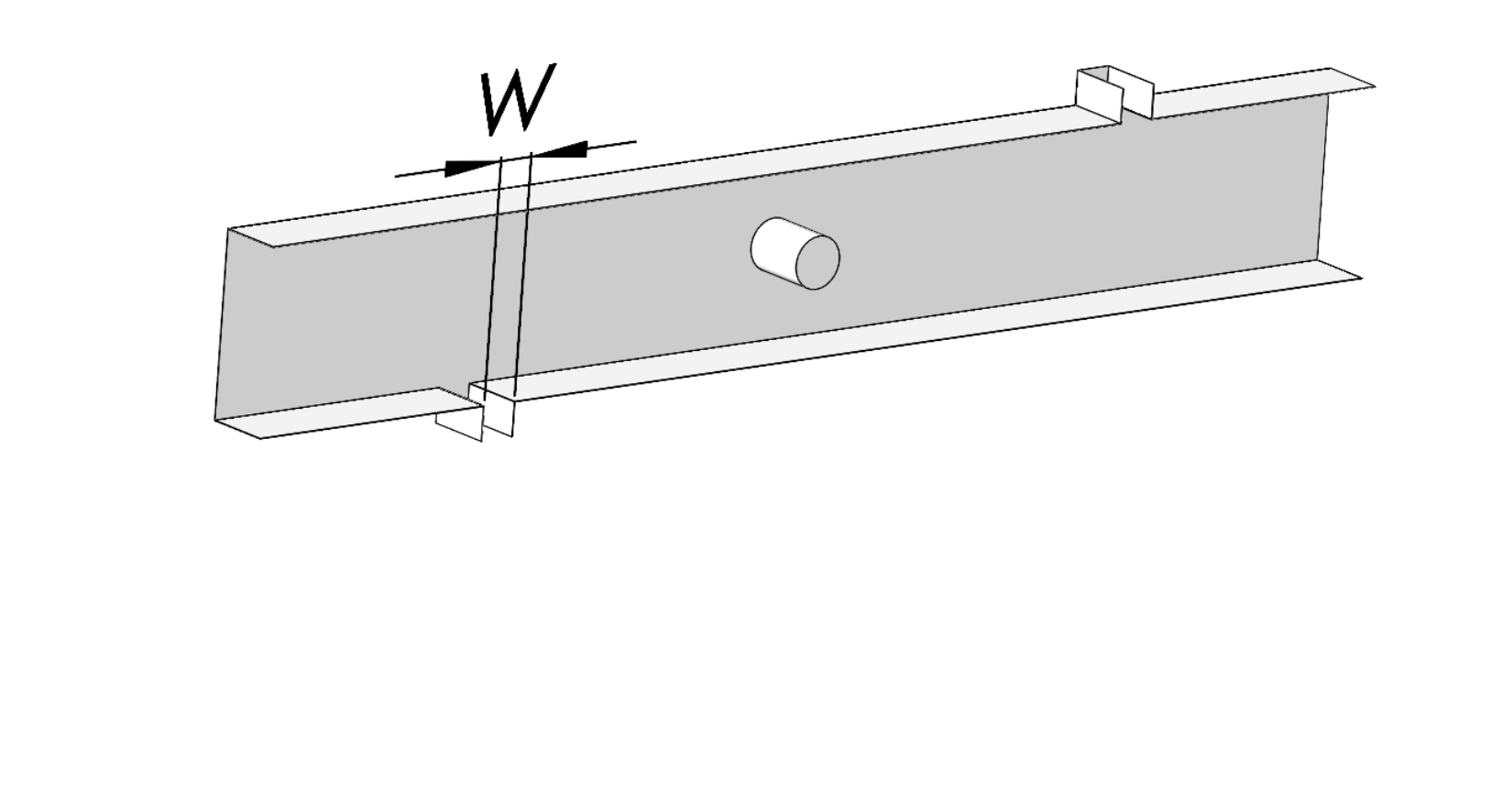}
    \end{minipage}
    \caption{Schematic representation of the (a) WTI and (b) NTI channels (flow is left to right).}
    \label{fig:channels}
\end{figure}

In this work, two types of channels were used, which differed only on pressure tap design (which were placed \mbox{$L=500$ \textmu m} up- and downstream from the cylinder face, on opposite walls): Narrow-Tap-Intake (NTI) and Wide-Tap-Intake (WTI) channels. A sketch of both channels can be seen in Figure \ref{fig:channels}.

The microfluidic devices were fabricated through standard PolyDiMethylSiloxane (PDMS) soft-lithography techniques from SU-8 coated silicon moulds, using a two-part Sylgard\textsuperscript{\textregistered} 184 PDMS polymer kit. The moulds were fabricated on single square pieces of silicon through optical lithography from aluminium coated Corning\textsuperscript{\textregistered} glass hard-masks, on which the channel designs were originally cut out through direct laser lithography. The PDMS was mixed thoroughly with the cross-linker in a 10:1 mass ratio and allowed to degas under vacuum in a desiccator for an hour. The mould's surface was made hydrophobic through a silanization process with TriMethylChloroSilane (TMCS), also for one hour, to aid in peeling the PDMS layer. The PDMS mixture was then poured into the mould and again allowed to degas under vacuum for another 30 minutes. The PDMS-filled mould was then cured at 80$^\circ$C in an oven for 30 minutes. The PDMS layer was then cut, punctured (inlet, outlet and pressure taps) and plasma-bonded to a clean glass microscope slide. 

\subsection{Flow control and dimensionless groups}
\label{sec:methods:flow control}

The fluids were driven into the channels with a neMESYS low-pressure syringe pump of 14:1 gear ratio (CETONI GmbH). A \mbox{5 ml} Hamilton\textsuperscript{\textregistered} Gastight syringe was used for most of the measurements (for lower-flow-rate measurements, namely for distilled water streak imaging, a 2.5 and a 0.1 \mbox{ml} syringes were employed), ensuring pulsation-free dosing. The syringe was connected to the inlet of the channel through flexible Tygon\textsuperscript{\textregistered} tubing and a stainless steel precision tip, Nordson Corporation. The outlet of the channel was left open. All existing air bubbles were removed from the pressure taps, which were sealed as pressure measurements were outside this work's scope.

Our flow regimes were defined by the Weissenberg number, \mbox{$100\leq Wi = \lambda U/R\leq500$}, from which we could compute the mean channel velocity, \mbox{$U$}, leading us to the flow rate, \mbox{$Q = 2L_\mathrm{c} H U$}, and to the Reynolds number, \mbox{$Re = \rho U R/\eta_0 < 1$}. Additionally, we computed the elasticity number (representing the ratio between elastic and inertial forces), \mbox{$El = Wi/Re = 620$ and $30323$}, for PEO4 and PEO8, respectively. The viscoelastic Mach number, \mbox{$Ma = \sqrt{WiRe}$}, was also computed.

\subsection{Micro-Particle Image Velocimetry}
\label{sec:methods:piv}

The flow kinematics were analysed through the velocity vector fields inside the microchannels, obtained through \textmu PIV (at room temperature). The fluids were seeded with \mbox{2 \textmu m} diameter, fluorescent tracing particles (Molecular
Probes\textsuperscript{\texttrademark} FluoSpheres\textsuperscript{\textregistered}; F8825 Nile red carboxylate-modified microspheres) in concentrations of approximately \mbox{0.5 wt\%} for both fluids. Additionally, sodium dodecyl sulphate (SDS) was added in a concentration of \mbox{0.1 wt\%} to avoid excessive particle adhesion to the PDMS walls or the microscope slide. The samples were filtered with a Thermo Scientific$^{\mathrm{TM}}$ Nalgene$^{\mathrm{TM}}$ syringe filter of \mbox{10 \textmu m} pore size, testing, beforehand, if the filtering process compromised the rheology of the samples.

The setup for the \textmu PIV measurements consisted of an inverted microscope (Leica Microsystems GmbH; DMI5000 M), equipped with a 20$\times$ magnification lens (Leica Microsystems GmbH; HCX PL Fluotar with a numerical aperture of \mbox{NA = 0.4}). A double-pulsed Nd:YAG laser system with \mbox{532 nm} wavelength was used to illuminate the whole fluid volume, and the images were taken with a CCD camera (Dantec Dynamics; FlowSense 4M MkII) of \mbox{2048$\times$2048 px} resolution at full frame (\mbox{1 px = 7.4 \textmu m}).

The measurements were done at the channel centre plane, which was identified by focusing the top and bottom of the channel and averaging the two. One hundred image pairs were taken with a trigger rate of \mbox{7.4 Hz}, synchronised with the laser pulses, and were averaged using a cross-correlation PIV algorithm (Dantec Dynamics; DynamicStudio v2.30) to create a single velocity vector field per measurement. The time between laser pulses was set independently for each measurement, depending on the flow rate. It was considered as the time that a particle would take to travel a fraction of an interrogation area (the whole measurement area was split into interrogation areas of \mbox{32$\times$32 px}) in the contractions, where the velocity is highest (the mean velocity in the contractions is \mbox{$U_\mathrm{contr} = U L_\mathrm{c}/(L_\mathrm{c}-R)$}). Two velocity vector fields were obtained for each flow regime, averaged and slightly smoothed using the \textit{smoothn} algorithm \citep{garcia2010,rodrigues2020,martinez2016}.

\subsection{Streak imaging}
\label{sec:methods:streak}

Flow pattern visualisations were also performed to qualitatively evaluate the instantaneous behaviour of the fluid flows. This was done by obtaining streak images, with long exposure times, of the fluorescent-particle-seeded fluid samples. The samples were prepared with \mbox{2 \textmu m} particles and SDS at the same concentrations and following the same procedure. 

The setup consisted of the same microscope now equipped with a 10$\times$ magnification lens (\mbox{NA $=0.3$}) and a \mbox{100 W} mercury lamp to illuminate the flow. The images were taken at the channels' centre planes with a CCD camera for fluorescence microscopy (Leica Microsystems GmbH; DFC350 FX 12-bit monochrome). Further details concerning the streak imaging and \textmu PIV setup can be found in Rodrigues et al.\citep{rodrigues2020}.

\section{Results and discussion}
\label{sec:results}

\subsection{Elastic weakly shear-thinning fluid}
\label{sec:results:PEO4}

\begin{figure}[htp]
    \centering
    \includegraphics[width=.8\textwidth]{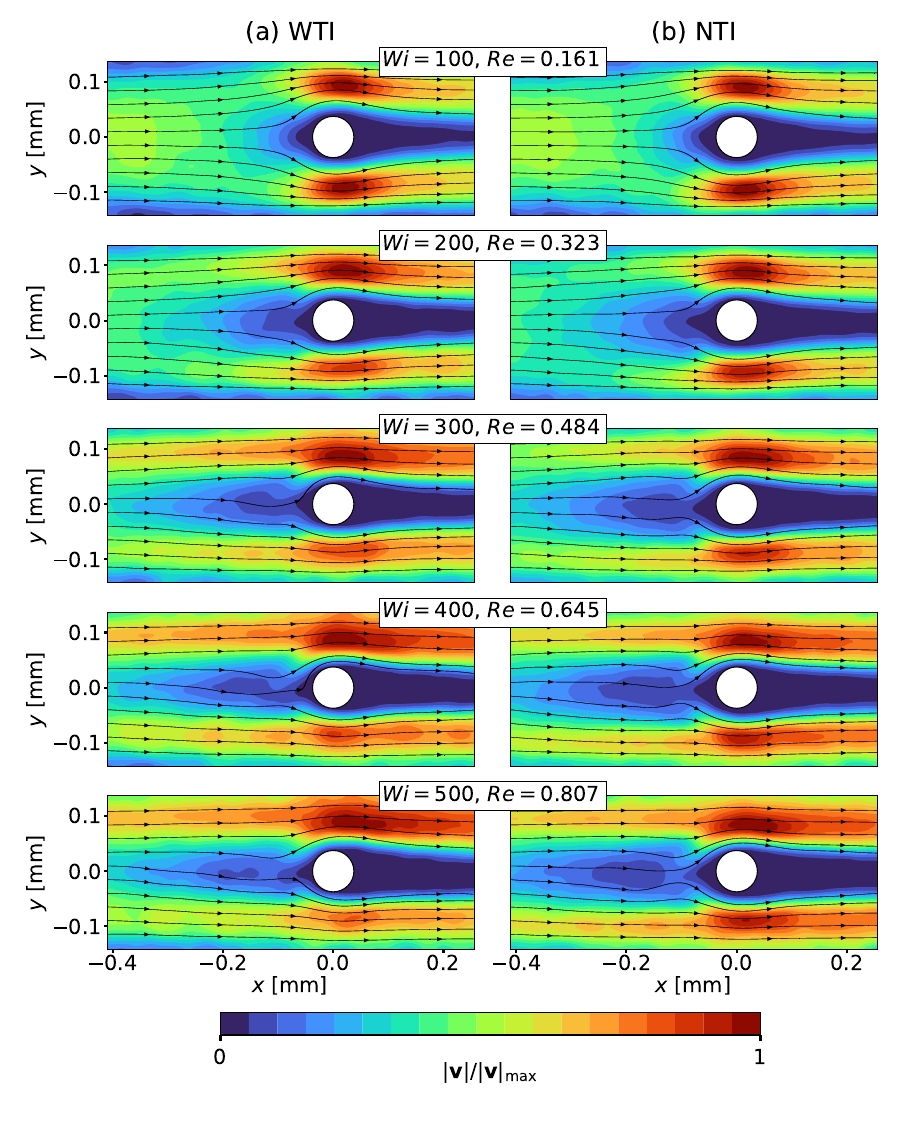}
    \caption{Time-averaged normalised velocity magnitude flow fields, $\lvert \mathbf{v} \rvert / \lvert \mathbf{v} \rvert_{\mathrm{max}}$, and superimposed streamlines obtained with PEO4 for (a) WTI and (b) NTI channels (flow is left to right).}
    \label{fig:velocity fields}
\end{figure}

Firstly, we shall discuss the results obtained with the elastic weakly shear-thinning fluid, PEO4. From the velocity vector fields obtained from the \textmu PIV measurements we computed the velocity magnitude fields, normalised with the maximum value found in each flow regime: $0\leq\lvert \mathbf{v}\rvert/\lvert\mathbf{v}\rvert_{\mathrm{max}}\leq 1$, where $\lvert\mathbf{v}\rvert=\sqrt{u^2+v^2}$. The results for each flow regime obtained for PEO4 can be seen for each pressure tap design, in Figure \ref{fig:velocity fields}.

For the initial flow regime of $Wi=100$, the flow is laterally symmetric for both channel designs. Fore-aft symmetry, however, is already broken at this stage as an extended downstream wake and a slight upstream wake, with possible forward stagnation point detachment, are already formed, consequences of fluid elasticity, similar to previous results\cite{nolan2016,haward2018,varchanis2020}. Increasing $Wi$ leads to differences between the two channel flows. Analysing the WTI flows (Figure \ref{fig:velocity fields}(a)), increasing $Wi$ to $200$ breaks the lateral symmetry as larger velocities can be seen in the positive-$y$ gap (the onset of lateral flow asymmetry is within a viscoelastic Mach number interval: $4<Ma<8$); this lateral asymmetry seems to remain, and even grow, while further increasing $Wi$. Notably, a slight bending of the upstream wake can be seen from $Wi$ of $300$ to $500$. The bending of the streamlines immediately upstream of the cylinder resembles the already-mentioned wineglass-shaped flow found for Boger fluids\citep{kenney2013,galindo2014,shi2015}, now in a half-wineglass shape in the positive-$y$ region. Lower velocities are found, for all $Wi$, at the wall containing the upstream pressure tap when compared to the opposing wall, and both up- and downstream wakes grow with $Wi$. Now, considering the NTI channel (Figure \ref{fig:velocity fields}(b)), the flow shows almost negligible lateral asymmetry for all $Wi$. For $Wi=300$ the full wineglass-shaped profile becomes visible, and intensifies with increasing $Wi$, as can be clearly seen from the bending of the streamlines upstream of the cylinder, similar to the results of Kenney et al.\cite{kenney2013}. Again, extended up- and downstream wakes are visible, growing with $Wi$, but the wall velocities are similar, not showing the discrepancies found with the WTI channel.

\begin{figure}[htp]
    \centering
    \includegraphics[width=.8\textwidth]{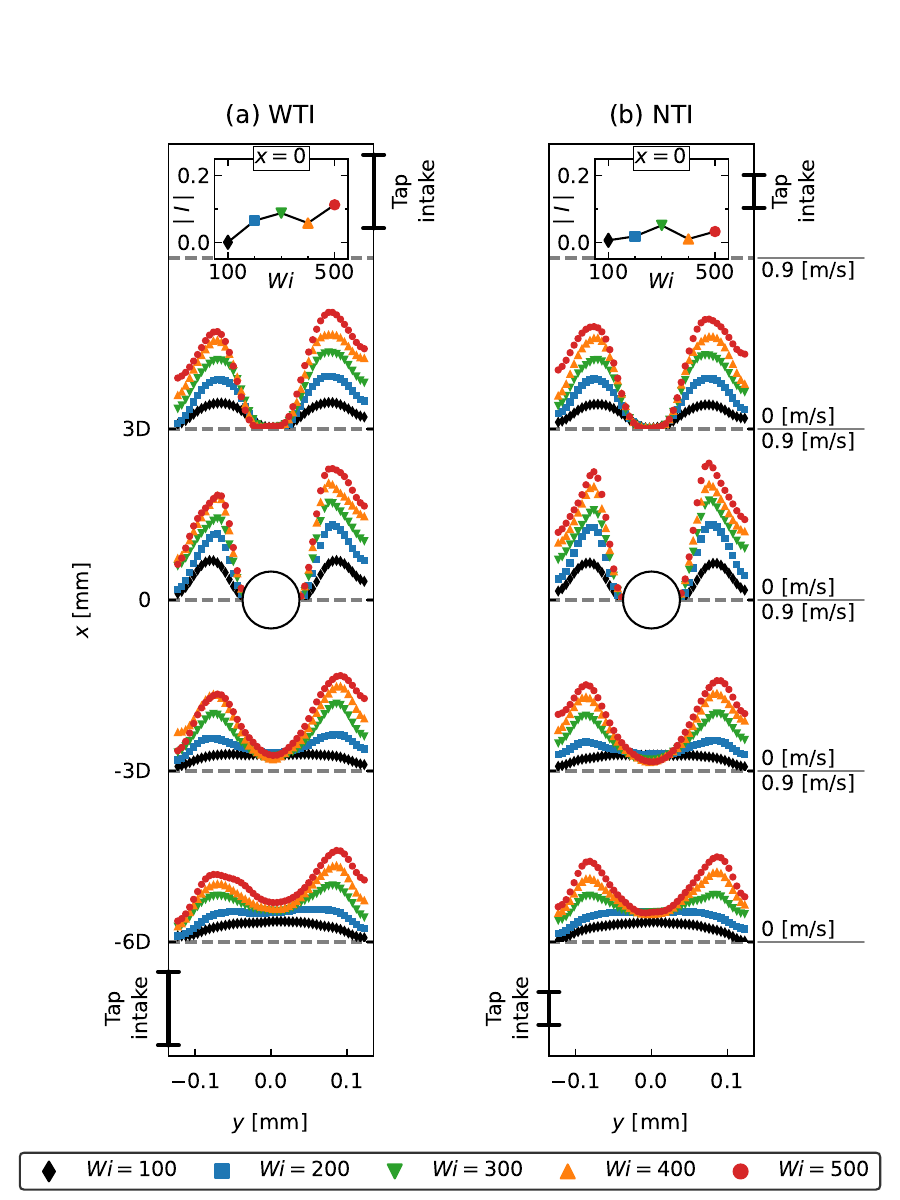}
    \caption{Time-averaged streamwise velocity profiles, $u$ [m/s], obtained at different axial positions (flow is bottom to top), and asymmetry parameter, $\lvert I\rvert$, for (a) WTI and (b) NTI channels for PEO4.}
    \label{fig:profiles PEO4}
\end{figure}

To gain insight into the flow's evolution throughout the channel, axial velocity profiles were plotted at several $x$-positions ($-6D$, $-3D$, $0$ and $3D$). Additionally, to better visualise the degree of lateral asymmetry in the flow, an asymmetry parameter was computed from the axial velocities found at the cylinder centre plane ($x=0$). This parameter was originally defined by Haward et al.\citep{haward2019}, comparing the streamwise velocities found at the centre of each gap surrounding the cylinder, as mentioned before in Section \ref{sec:intro}. However, as the maximum gap velocities are not always found at their mid-points, we chose to modify the asymmetry parameter: \mbox{$I = (u^+-u^-) /(u^++u^-)$}, where $u^-$ and $u^+$ are the maximum velocities found in the negative-$y$ and the positive-$y$ gaps, respectively. Thus, positive and negative $I$ values correspond, respectively, to fluid flowing preferentially through the positive and negative-$y$ gaps, while \mbox{$I = 0$} means perfect lateral symmetry. The streamwise velocity profiles along the channel and the asymmetry parameters, computed for each flow regime, for both channels, can be seen in Figure \ref{fig:profiles PEO4}.

Firstly, generally observing the streamwise velocity profiles upstream of the cylinder, the growth of the upstream wake with $Wi$ is clear. For \mbox{$Wi = 100$} the profiles exhibit the classic undisturbed fully-developed flow shape and increasing the flow rate leads to a relative drop in the centre-channel velocity (\mbox{$y = 0$}), gaining relevance as we approach the forward stagnation point. The downstream wake is also clear as three cylinder diameters downstream, the streamwise velocities near the centre of the channel are practically null. The velocity profiles and asymmetry parameters reinforce the observations made from the previously discussed velocity magnitude flow fields (Figure \ref{fig:velocity fields}), with the WTI profiles showing the lateral flow asymmetry throughout the entire area of observation once it onset at \mbox{$Wi = 200$}. From then on, the asymmetry parameter remains relatively constant, with a maximum value of $I\approx0.11$ for $Wi=500$. For the NTI channel, there is practically no lateral flow asymmetry for all tested $Wi$.

As mentioned previously, the necessary conditions for the onset of the reported lateral flow asymmetry were found to be fluid elasticity and shear-thinning\cite{haward2020,varchanis2020}. Thus, it was found pertinent to evaluate possible viscosity variation effects in the flow of both fluids. We were able to compute the local viscosities for all flow regimes through the C--Y models (equation \ref{eq:CY}), using the magnitude of the local shear rates, $\dot\gamma$, as input. The local shear rates were calculated through the following expression \citep{bird1987dynamics}:
\begin{equation}
    \dot\gamma = \sqrt{(1/2)(\mathbf{D}:\mathbf{D})}
    = \sqrt{(1/2) \sum \mathbf{D}_{ij}^2}\,,
    \label{eq:deformation}
\end{equation}
where, $\mathbf{D}=(\nabla \mathbf{v}+\nabla\mathbf{v}^\mathrm{T})/2$, is the deformation rate tensor. The shear rate range for both fluids was computed as the minimum and maximum local values found in all flow regimes and can be seen, plotted as vertical bands, in Figure \ref{fig:useful}. For PEO4, the shear rate ranged: \mbox{$5.3<\dot\gamma_\mathrm{PEO4}<10776.7$ s$^{-1}$}, which induced a small viscosity variation: \mbox{$12.91<\eta_\mathrm{PEO4}<25.24$ mPa$\cdot$s}. This small viscosity range was expected, regardless of the shear rate range, due to PEO4's near-Boger behaviour and, thus, the viscosity variation's role in PEO4's lateral asymmetry, and overall flow behaviour, can be disregarded.

\begin{figure}[htp]
    \centering
    \includegraphics[width=.8\textwidth]{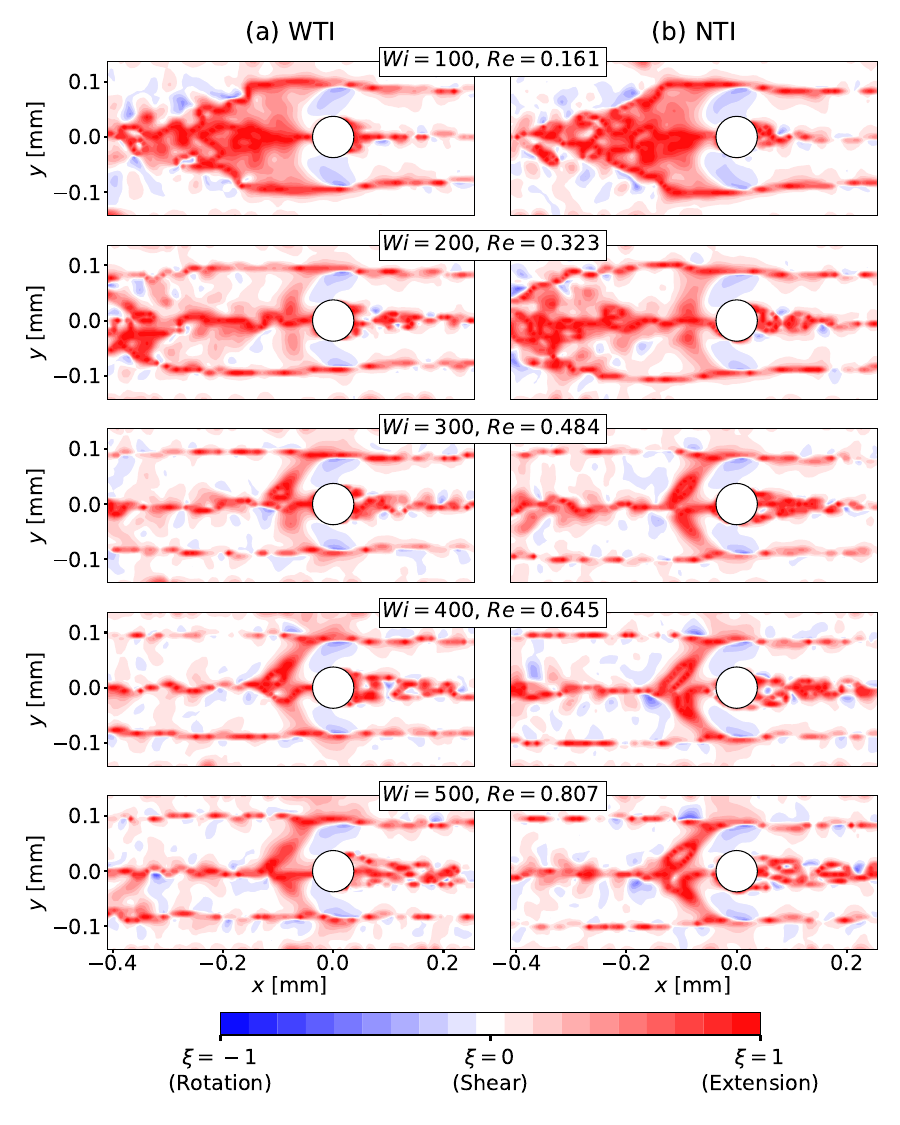}
    \caption{Time-averaged flow-type parameter flow fields, $\xi$, computed for (a) WTI and (b) NTI channels for PEO4 (flow is left to right).}
    \label{fig:flowtype PEO4}
\end{figure}

Some additional insight regarding the flow can be gathered through the flow-type parameter, $\xi$, which gives information concerning the contributions of rotation, shear and extension to the flow. It can be locally defined through the magnitudes of the aforementioned deformation rate tensor, $\mathrm{\mathbf{D}}$, and the vorticity tensor, $\mathrm{\mathbf{\Omega}} = (\nabla \mathbf{v}-\nabla\mathbf{v}^\mathrm{T})/2$, according to the following expression\citep{nolan2016,rodrigues2020}:

\begin{equation}
    \xi = \frac{\lvert \mathbf{D}\rvert - \lvert \boldsymbol{\Omega}\rvert}{\lvert \mathbf{D}\rvert + \lvert \boldsymbol{\Omega}\rvert}\,.
    \label{eq:flowtype}
\end{equation}
Flow-type parameters of $-1$, $0$ and $1$ correspond, respectively, to pure rotational, pure shear and pure extensional flows. Flow-type parameter flow fields were plotted for each channel and can be seen for PEO4 in Figure \ref{fig:flowtype PEO4}.

For $Wi=100$, a significant area of extensional flow can be seen upstream of the cylinder, being progressively replaced with shear flow as $Wi$ increases and the upstream wake is formed and grows. Notably, three streams of extensional flow can be seen both up- and downstream of the cylinder for all $Wi$, similarly to the reports of previous works\cite{nolan2016,rodrigues2020,haward2018}. The streams located at the channels' centre are related to the compression of the fluid moving towards the forward stagnation point and the high extensional stresses associated with the downstream stagnation point, and the two remaining streams stand as a result of the stretching of the fluid as it flows into and out of the gaps. In both gaps, near the cylinder face, areas of rotational flow can be seen. Looking now specifically at the WTI channel flow-type parameter fields (Figure \ref{fig:flowtype PEO4}(a)), as the flow asymmetry onsets and grows, we can see the strengthening of an asymmetric extensional flow region near the forward stagnation point, which seems to shift upstream slightly. This extensional flow area represents the fluid diverging mainly into the positive-$y$ gap. With the NTI channel, this extensional flow region near the forward stagnation point can also be seen, and, despite displaying similar strength into either gap, its shape is complex and shows a degree of asymmetry. The axial movement of these extensional flow regions observed with both channels could be related to the detachment of the cylinder's forward stagnation point.

\begin{figure}[htp]
    \centering
    \includegraphics[width=.75\textwidth]{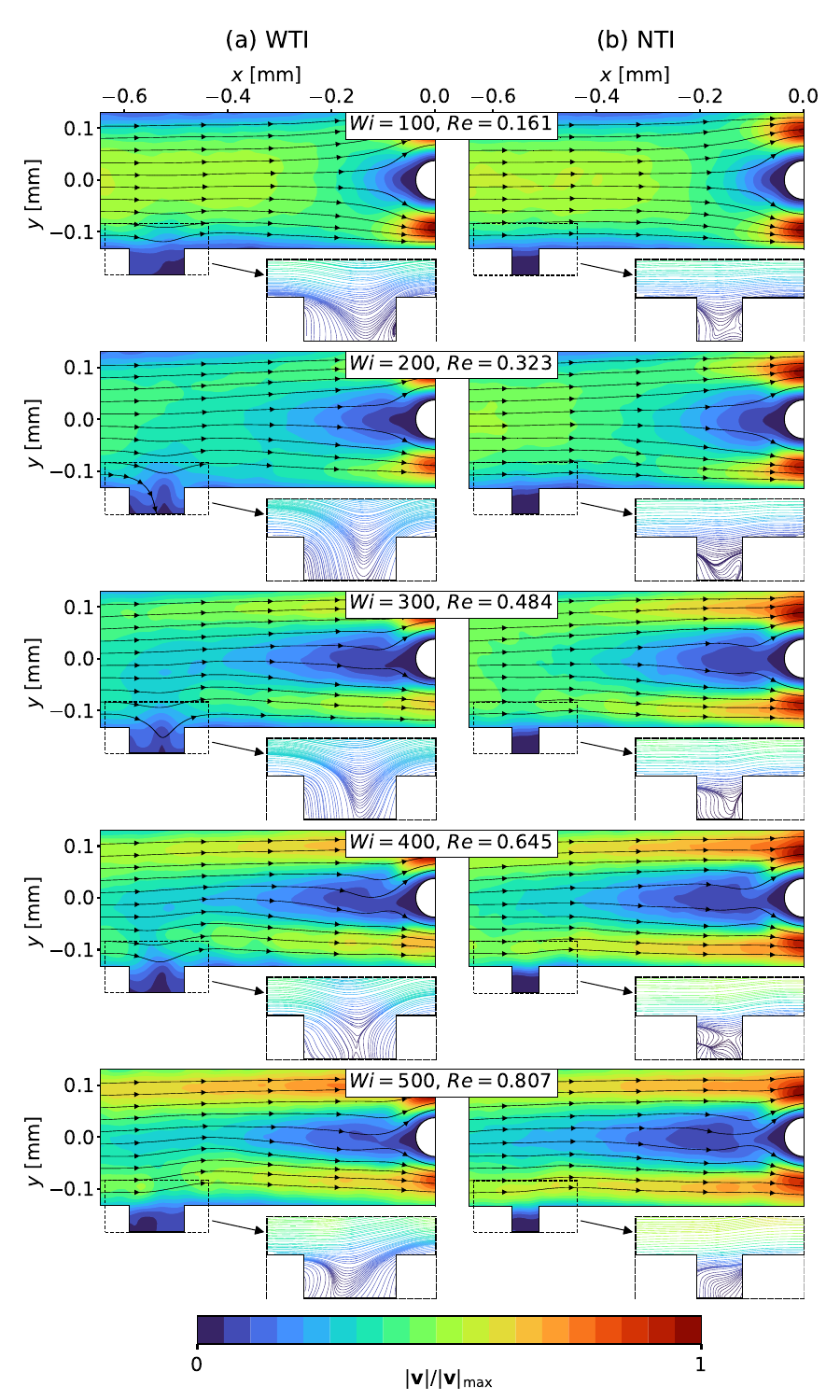}
    \caption{Time-averaged velocity magnitude flow fields, $\lvert \mathbf{v} \rvert/\lvert \mathbf{v} \rvert_\mathrm{max}$, and superimposed streamlines obtained with PEO4 for (a) WTI and (b) NTI channels, past the upstream pressure tap intake (flow is left to right).}
    \label{fig:Taps insets PEO4}
\end{figure}

As could be gathered from the analysed data (Figures \ref{fig:velocity fields}--\ref{fig:flowtype PEO4}), there are significant differences between the flow in the two channels. It seems that the wider pressure tap intake influences the flow to such a degree that it may lead to lateral flow asymmetry, forcing the fluid into the opposing gap, while the narrow intake seems not to have such a noticeable influence, as flow symmetry was practically maintained for the entirety of the tested flow regimes. In order to visualise the upstream pressure taps, additional measurements were conducted, shifting the measurement area slightly upstream. The new velocity magnitude fields obtained for PEO4, including the tap intakes, can be seen in Figure \ref{fig:Taps insets PEO4}. The plots also show a zoomed inset of the tap intake.

Viscoelastic fluids are susceptible to flow instabilities and perturbations, mainly when employed on small length scale channels. At the tap intakes, we have essentially \textit{flow past a cavity}, where our ``cavity'' is deep and of a relatively complex shape (the pressure tap). Generally speaking, within the cavity, two flow regions can be observed, firstly a recirculating region (with one or more vortices) deeper within the hole and, secondly, the main flow, which, as it encounters the geometrical broadening of the channel, protrudes into it. The two regions are separated by the dividing streamline, i.e., the last streamline that penetrates into the cavity and bends to rejoin the main flow. Cochrane et al.\citep{cochrane1981} and Kim et al.\citep{kim2000} showed experimental and numerical results for the flow of Boger and Newtonian fluids past different sized macroscopic cavities, reporting several interesting behaviours. Firstly, fluid inertia and elasticity both lead to streamline asymmetry within the cavity, shifting the location of the recirculating vortex, but in opposite directions and can counteract each other\citep{cochrane1981}. Moreover, fluid elasticity seems to shift the recirculating vortex and the dividing streamline deeper into the cavity\citep{cochrane1981}. The geometry of the cavity also greatly influences the flow behaviour and deeper holes can even lead to multiple vortices\citep{cochrane1981}. Lastly, Kim et al.\citep{kim2000} found a set of flow transitions within the cavity at very low Reynolds numbers, essentially decoupling elastic effects from inertia. The authors found that the initial symmetry of the recirculating area was lost and the dividing streamline progressed deeper into the cavity as the flow rate increased, until the flow became time-dependent, with the dividing streamline oscillating within the cavity, and the recirculating vortex subdividing into several counter-rotating vortices in the neutral flow direction. The authors also encountered time-dependent flow up- and downstream of the cavity, captured through pressure measurements, and mentioned the possibility of the flow becoming unstable at large Deborah numbers (representing the ratio of relaxation time to time of observation --- another measure of elastic effects). They additionally consider the possibility that the flow transitions and time-dependence within the cavity propagated through elastic effects up- and downstream.

Our larger intake (WTI) allows the dividing streamline to be located deeper within the tap. There also appears to be a transition of flow shapes inside the intake between flow regimes, all showing a degree of asymmetry and strong streamline bending (which agrees with the shapes shown by Cochrane et al.\cite{cochrane1981}). For the NTI flows, the bending of the streamlines is much more subtle and the dividing streamline more shallow. We can gather that the intakes seem to be causing the flow to ``trip'', slowing it down near the wall containing the upstream pressure tap (negative-$y$), and bending the flow (seen from the streamline curvature near the intakes). This is true for both tap designs, but much more notorious for the WTI, which further hints at the possibility that the flow asymmetry may be the result of a flow perturbation induced by the wider intake. Moreover, similarly to what was done before, we computed the local flow-type parameter to evaluate the flow near the intake. These plots are not shown here for conciseness but are given as supplementary material and we shall briefly discuss them here. Starting with the WTI flows, large extensional flow regions are seen connecting the inside of the pressure tap to the upstream extensional flow streams. Moreover, strong rotational flow can be found immediately up- and downstream of the intake, near its correspondent channel wall. Both regions are representative of the fluid's penetration into the pressure tap intake, bending the streamlines and stretching the fluid in and out of it. Accordingly, both the extensional and rotational flow regions are stronger the deeper the fluid penetrates into the tap. With the NTI channel, similar regions are visible but much less pronounced than with the WTI.

\begin{figure}[htp]
\centering
\begin{minipage}[t]{.40\textwidth}
    \centering 
    PEO4
    \\
    \footnotesize{($Wi = 300$, $Re = 0.484$)}
    \\
    \footnotesize{(a) WTI}
    \includegraphics[width=\linewidth]{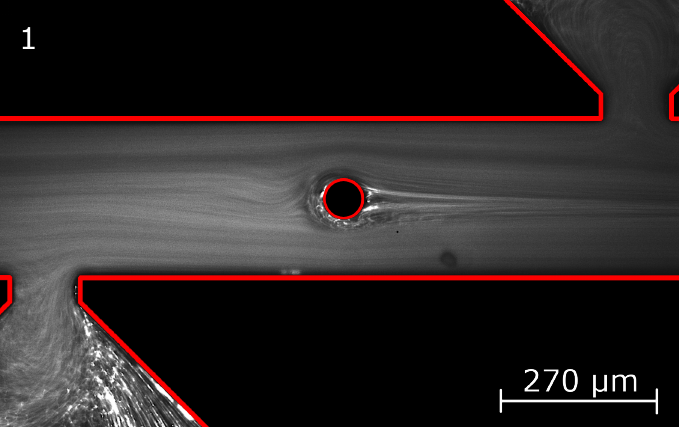}
    \includegraphics[width=\linewidth]{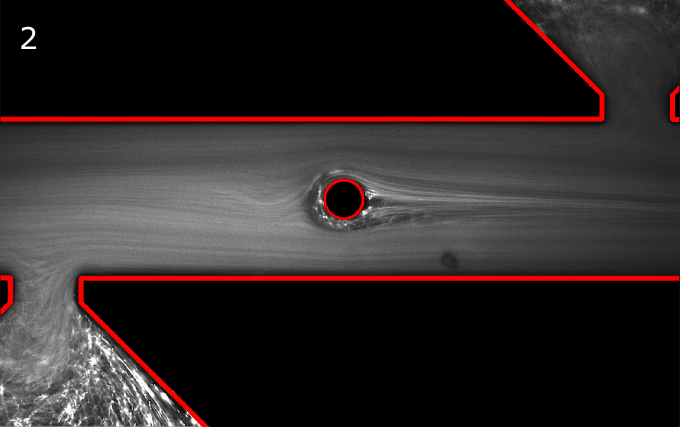}
    \footnotesize{(c) NTI}
    \includegraphics[width=\linewidth]{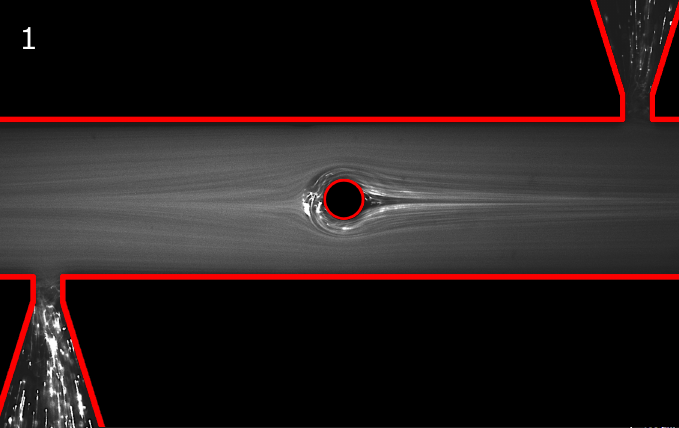}
    \includegraphics[width=\linewidth]{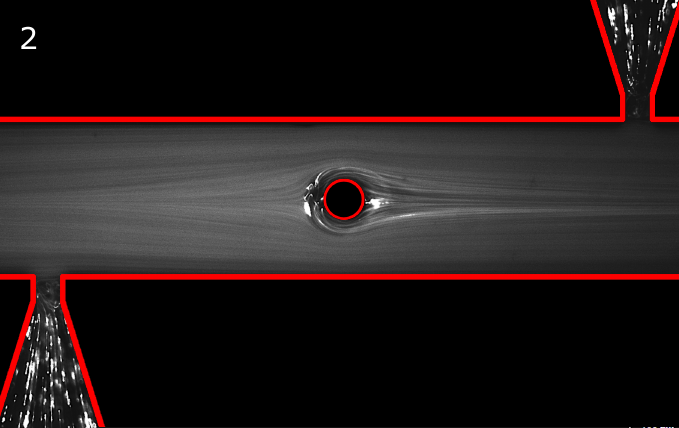}
\end{minipage}
\begin{minipage}[t]{.40\textwidth}
  \centering
    Water
    \\
    \footnotesize{($Wi = 0$, $Re = 0.484$)}
    \\
    \footnotesize{(b) WTI}
    \includegraphics[width=\linewidth]{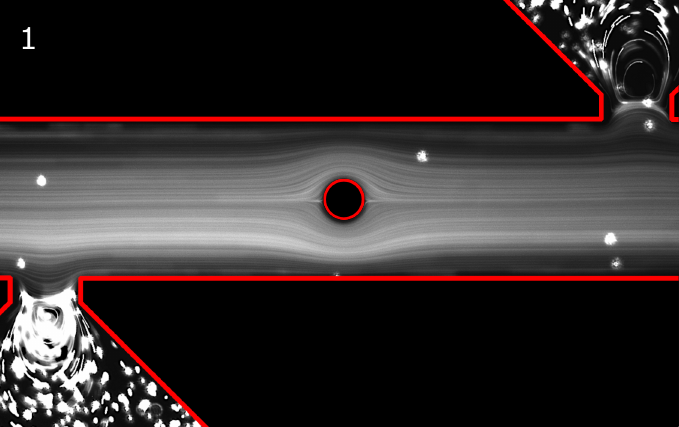}
    \includegraphics[width=\linewidth]{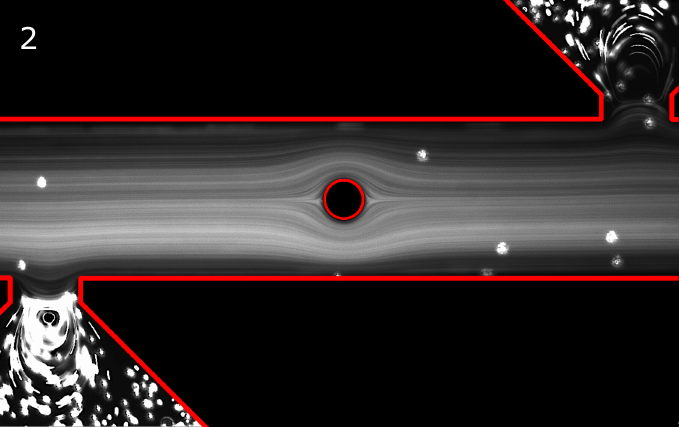}
    \footnotesize{(d) NTI}
    \includegraphics[width=\linewidth]{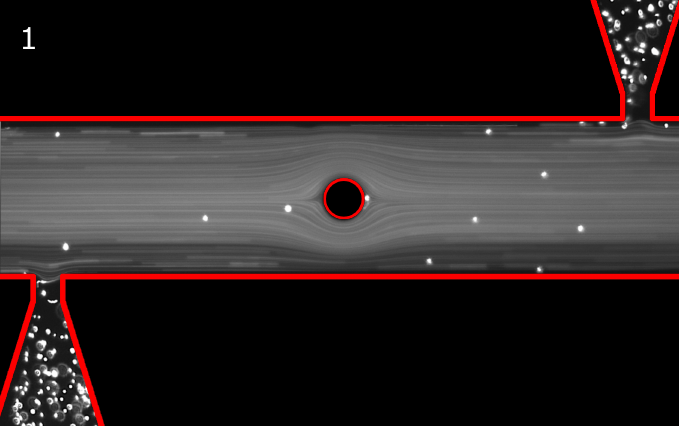}
    \includegraphics[width=\linewidth]{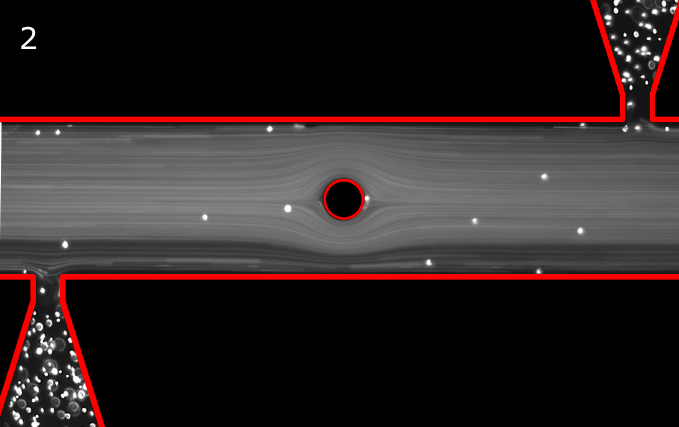}
\end{minipage}
\caption{Pairs of consecutive streak images (1 and 2), obtained with (left) PEO4 and with (right) Water, on ((a) and (b)) WTI and ((c) and (d)) NTI channels (flow is left to right).}
\label{fig:Streak PEO4}
\end{figure}

Previous reports of unsteady flow in the studies of viscoelastic instabilities\cite{kenney2013,nolan2016,zhao2016,haward2019,varchanis2020,hopkins2022} and flow past cavities\cite{kim2000} lead to the necessity of evaluating the instantaneous flow patterns; thus, flow visualisations were conducted, taking two consecutive streak images for each channel design for $Wi = 300$. Additionally, flow visualisations were also conducted with distilled water to evaluate the role of elasticity, serving as a Newtonian control, at the same Reynolds number. The streak images obtained for both fluids (PEO4 and water), in both channels, can be seen in Figure \ref{fig:Streak PEO4}.

Looking at the results obtained with the WTI channel, the Newtonian flows (Figure \ref{fig:Streak PEO4}(b)) remain symmetric and steady, while for PEO4 (Figure \ref{fig:Streak PEO4}(a)), upstream of the cylinder, the bending of the streamlines and the stagnation point´s negative-$y$ shift is striking, as well as its slight detachment from the cylinder face, all corroborating the \textmu PIV results previously discussed. Notably, the streamlines begin to slightly bend far upstream of the cylinder (and upstream of the half-wineglass ``pinch''), due to the upstream perturbation caused by the pressure tap. Furthermore, the pressure tap intakes show quite deep penetration and asymmetry of the dividing streamline, particularly the upstream tap intake, similar to the deep hole results obtained by Cochrane et al.\citep{cochrane1981}. Also, streamline crossing can be seen, which suggests three-dimensional flow. There is little difference between the two PEO4 images except for a slight modification of the dividing streamline position inside the pressure tap, suggesting that the flow is practically steady. However, from our visualisations, we gather that, within the taps, there is significant time-dependency with the recirculating vortex shifting up and down the intake, similar to the results of Kim et al.\cite{kim2000}. We have also gathered that the deeper the streamline penetrates into the intake (related to the intake geometry and fluid elasticity\cite{cochrane1981}), the larger the lateral asymmetry becomes. The further penetration of the streamline is correlated to a larger deceleration of the fluid near the pressure tap, which could be sufficient for a flow rate imbalance between the gaps, depending on fluid elasticity. The strong stresses induced on the fluid by its flow into the intake (which could be related to the depth that it reaches) should require a relaxation of the polymer molecules as it rejoins the main flow, decelerating it locally, similarly to the theory presented by Haward et al.\cite{haward2018} for their extremely long downstream wakes. For the NTI channel (Figure \ref{fig:Streak PEO4}(c)), there is less significant bending of the streamlines near the forward stagnation point (which has again detached from the cylinder face) and there is no noticeable lateral asymmetry. However, the transient motion of the dividing streamline and evidence of three-dimensional flow is still noticeable at the upstream tap intake. Very slight vertical drifting of the forward stagnation point could also be observed, linked to the penetration of the dividing streamline into the tap intake. The Newtonian fluid (Figure \ref{fig:Streak PEO4}(d)) showed the same steady, symmetric flows as with the WTI channel. In all PEO4 streak images shown, the forward stagnation point's location seems to belong to a flow structure that surrounds the cylinder, reminding us of the ``jacket of highly stressed, locally viscosified fluid'' reported by Haward et al. \cite{haward2018}(p. 30), which provokes an apparent enlargement of the cylinder radius, detaching the stagnation point. 

The fact that lateral flow asymmetry is found only with the WTI channels, and that it seems to be connected to the penetration of the dividing streamline into the pressure tap intake, leads us to believe that the phenomenon is due, not to a viscoelastic instability similar to previous reports\cite{haward2019,haward2020,nolan2016,ribeiro2014,zhao2016,varchanis2020,hopkins2022}, but due to a flow perturbation caused by the pressure taps and fluid elasticity. Furthermore, no evidence of lateral flow asymmetry was found experimentally by Varchanis et al.\cite{varchanis2020}, who tested the same fluid, nor encountered significantly in any other study with non-shear-thinning elastic fluids, which further corroborates our hypothesis. The fact that Varchanis et al.\cite{varchanis2020} also did not find any evidence of the characteristic streamline ``pinch'' we found with our NTI channel could be explained by our much smaller channel aspect ratio ($\alpha=0.37\ll5$), as previous studies have shown that increasing channel confinement anticipates the wineglass instability's onset\cite{kenney2013}.

It is worth highlighting that pressure taps, such as the ones used, may introduce errors in the pressure measurements. As the fluid flows past the hole, normal stresses bend the streamlines, and the generated tensile stresses lead to lesser pressures in the cavity compared to the bulk flow\citep{higashitani1975}. Thus, the measured pressure values would have an associated negative error. Our pressure tap configuration is set for differential pressure measurements, allowing us to evaluate the pressure drop provoked by the presence of the cylinder. When the local shear rate is similar at both tap intakes, the hole-pressure error would be equally present in both up- and downstream taps, thus cancelling itself on the differential value\citep{scott2004}. However, when such is not the case, the differences in the hole pressures at either tap would be an interesting research subject in itself, as it is related to the asymmetry of the flow and should be taken into account when conducting pressure measurements.

\subsection{Elastic shear-thinning fluid}
\label{sec:results:PEO8}

\begin{figure}[htp]
    \centering
    \includegraphics[width=.8\textwidth]{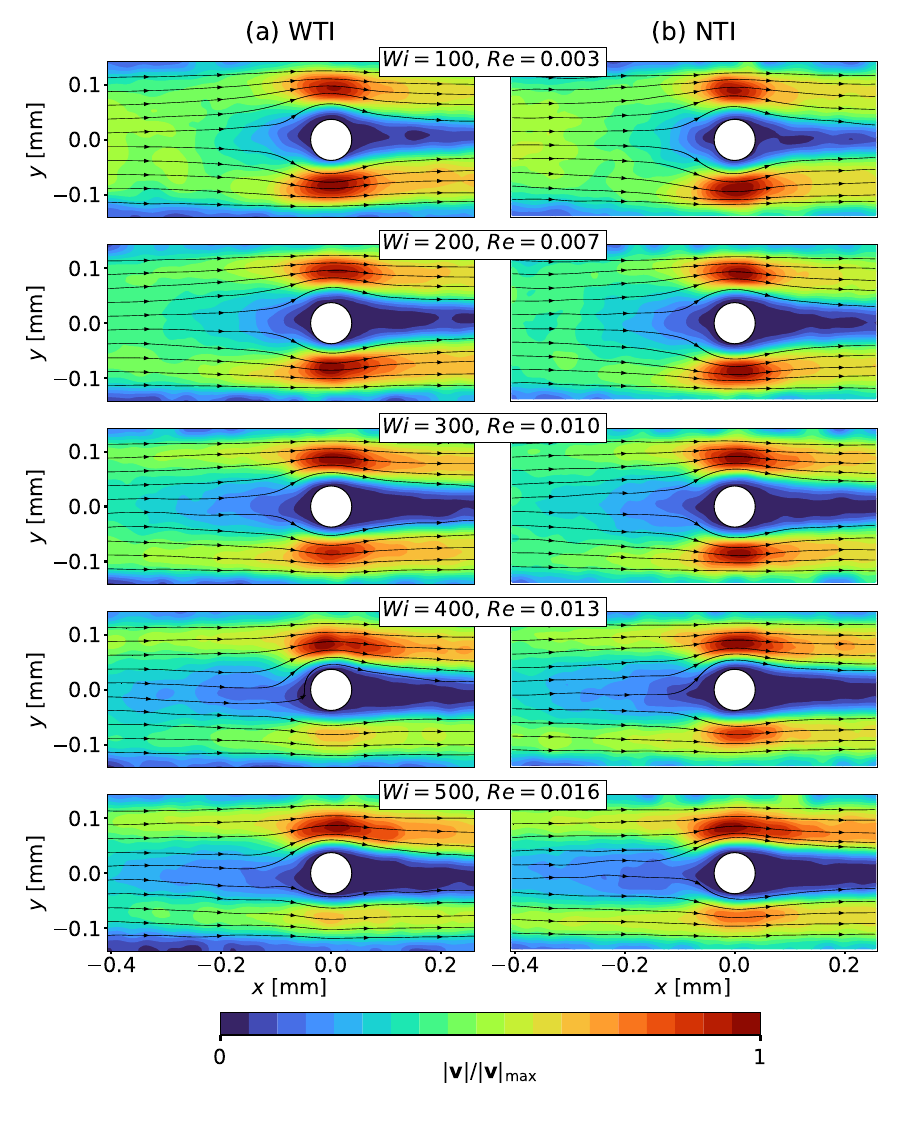}
    \caption{Time-averaged normalised velocity magnitude flow fields, $\lvert \mathbf{v} \rvert / \lvert \mathbf{v} \rvert_{\mathrm{max}}$, and superimposed streamlines obtained with PEO8 for (a) WTI and (b) NTI channels (flow is left to right).}
    \label{fig:velocity fields PEO8}
\end{figure}

In this section we shall go over all the results obtained with the elastic shear-thinning fluid, PEO8, analogously to the previous section for PEO4. In Figure \ref{fig:velocity fields PEO8} are shown the normalised local velocity magnitude flow fields, $\lvert \mathbf{v} \rvert / \lvert \mathbf{v} \rvert_{\mathrm{max}}$, obtained for both channels with PEO8.

PEO8's velocity fields are very similar for either tap design. For $Wi=100$, there is no fore-aft symmetry, with an extended downstream wake and a smaller upstream wake already observable, and both grow with $Wi$, possibly with stagnation point detachment from the cylinder face, just as was observed with PEO4 and previous works on elastic fluids. For $Wi=400$ a slight bending of the upstream wake, particularly for the wider intake, is noted; however, the bending of the streamlines near the forward stagnation point is not as pronounced as for PEO4, showing no evidence of the wineglass or half-wineglass shape. Lateral flow asymmetry, with fluid flowing preferentially through the positive-$y$ gap, onset for both WTI and NTI channels, contrarily to PEO4 which was only found to such a degree with the WTI. Furthermore, for the larger tap intake (Figure \ref{fig:velocity fields PEO8}(a)) the lateral flow asymmetry seems to onset earlier ($200<Wi<300$, $1.1<Ma<1.7$) than for the NTI ($300<Wi<400$, $1.7<Ma<2.3$) and lower velocities are found at the wall containing the upstream pressure tap. 

\begin{figure}[htp]
    \centering
    \includegraphics[width=.8\textwidth]{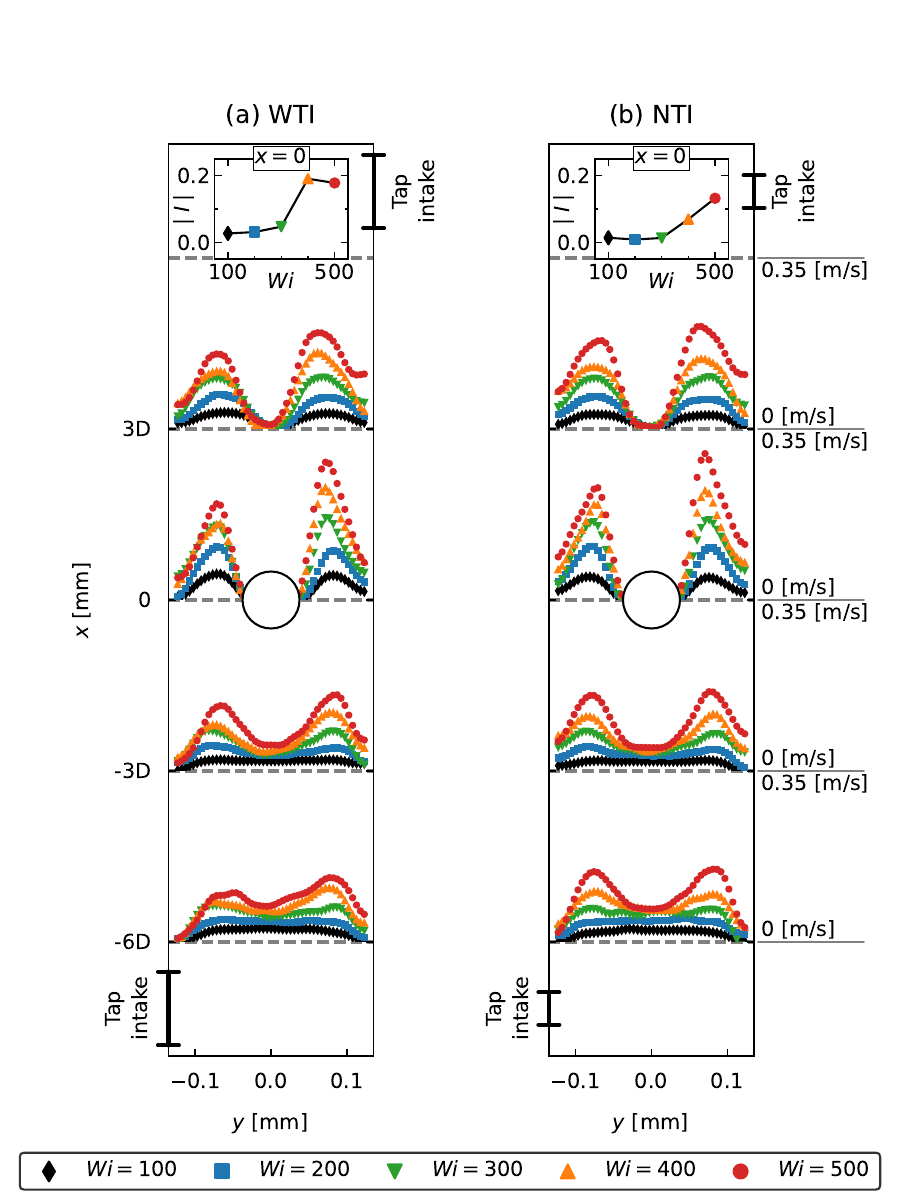}
    \caption{Time-averaged streamwise velocity profiles, $u$ [m/s], obtained at different axial positions (flow is bottom to top), and asymmetry parameter, $\lvert I\rvert$, for (a) WTI and (b) NTI channels for PEO8.}
    \label{fig:profiles PEO8}
\end{figure}

Again, the streamwise velocity profiles, $u$, along the channels and the asymmetry parameters, $I$, were computed and are shown, for PEO8, in Figure \ref{fig:profiles PEO8}. For $Wi=100$, upstream of the cylinder, the streamwise velocity profiles show the usual fully developed shape. Increasing $Wi$ again leads to a decrease in relative velocity at the centre of the channel ($y=0$), related to the upstream wake, however, with PEO8, this decrease in channel-centre velocity is not as significant as for PEO4. This shows that the upstream wake is not as strong for the shear-thinning fluid. Moreover, with the WTI channel, once lateral flow asymmetry onset, it became apparent upstream of the cylinder, after the pressure tap intake, with lower velocities at the negative-$y$ portion of the channel (again similar to the results obtained with PEO4), whereas the NTI flow seems symmetric up until it reaches the cylinder. Considering the asymmetry parameter plots, for the WTI flows (Figure \ref{fig:profiles PEO8}(a)), asymmetry onset between $Wi=200$ and $300$, increasing notably to an asymmetry parameter $I\approx0.19$ for $Wi=400$, seemingly stabilising thereafter. For the NTI flows, the lateral flow asymmetry increases steadily between its onset and the most elastic tested flow, $Wi=500$, where $I\approx0.13$.

\begin{figure}[htp]
    \centering
    \includegraphics[width=.8\textwidth]{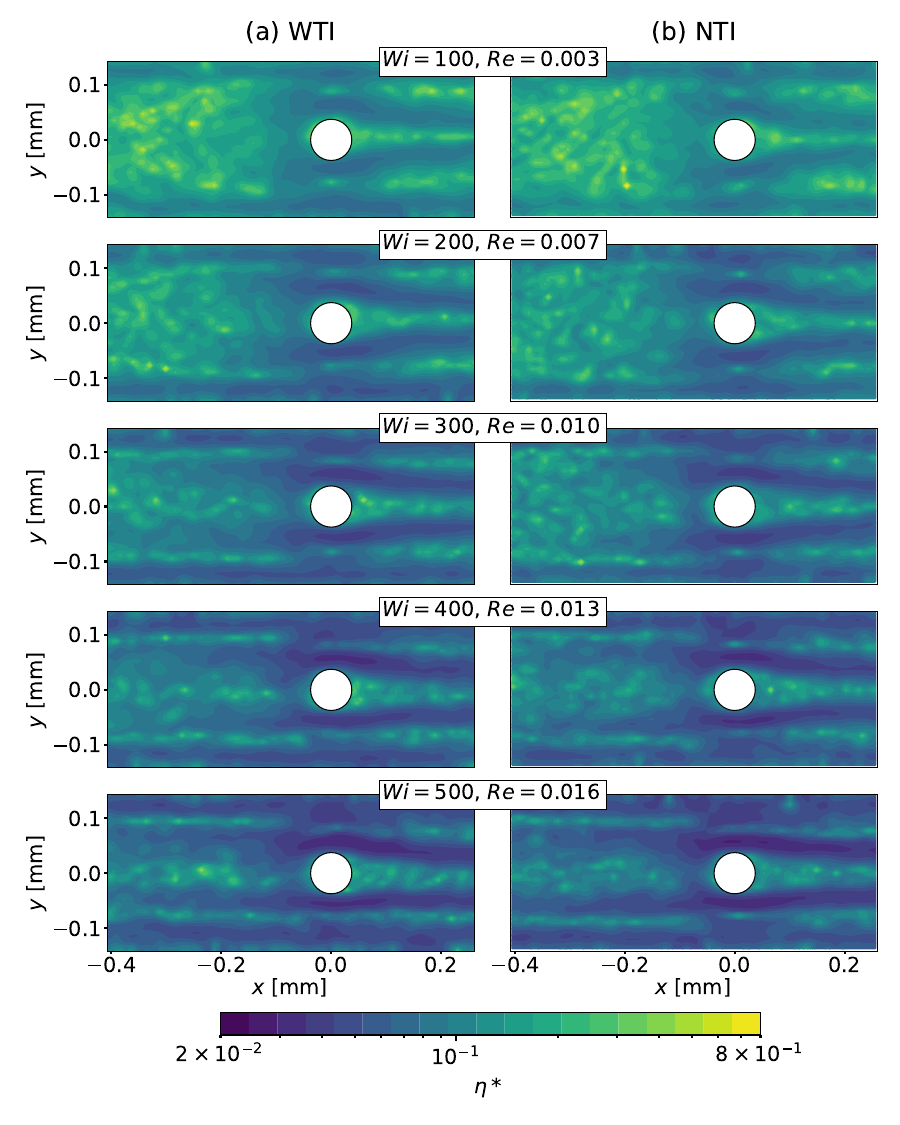}
    \caption{Time-averaged normalised viscosity flow fields, $\eta^*$, obtained for (a) WTI and (b) NTI channels for PEO8 (flow is left to right).}
    \label{fig:viscosity PEO8}
\end{figure}

Being PEO8 a strongly shear-thinning fluid, the viscosity variations within the flow are expected to be significant; the maximum viscosity variation found for PEO8 was: \mbox{$9.04<\eta_\mathrm{PEO8}<210.50$} mPa$\cdot$s, which corresponds to a shear rate range: \mbox{$2.2<\dot\gamma_\mathrm{PEO8}<4383.7$} s$^{-1}$. This viscosity variation, $\Delta \eta_\mathrm{PEO8} \approx 200$ mPa$\cdot$s, is indeed much more significant than the range achieved by the weakly shear-thinning PEO4: $\Delta \eta_\mathrm{PEO4} \approx 12$ mPa$\cdot$s. In order to more easily visualise the viscosity variations, it was found useful to normalise the local viscosities similarly to the work of Nolan et al.\citep{nolan2016}: \mbox{$\eta^* = (\eta(\dot\gamma) - \eta_\infty)/(\eta_0 - \eta_\infty)$}, where $\eta^*$ varies between $0$ (\mbox{$\eta(\dot\gamma) = \eta_\infty$}) and $1$ (\mbox{$\eta(\dot\gamma) = \eta_0$}). It is necessary, however, to keep in mind that this normalisation scales the viscosity variations within the two Newtonian plateaus, meaning that large variations in $\eta^*$ do not necessarily correspond to large true viscosity differences, particularly if the fluid has a reduced degree of shear-thinning. For PEO8, with this normalisation scheme, the effective viscosity range (\mbox{$9.04<\eta_\mathrm{PEO8}<210.50$} mPa$\cdot$s) is correspondent to a normalised viscosity range: $0.029<\eta^*<0.744$. PEO8's local viscosity flow fields are shown in Figure \ref{fig:viscosity PEO8}. PEO4's local viscosity fields are given as supplementary material.

The normalised local viscosity flow fields (Figure \ref{fig:viscosity PEO8}) were plotted with a logarithmic-scaled colour bar, meaning the contrasts between bright colours (right-hand side of the colour bar) represent larger viscosity variations than darker ones. For the initial regime shown in Figure \ref{fig:viscosity PEO8}, $Wi=100$, upstream of the cylinder the flow resembles slug flow, with a high viscosity, faster moving central area lubricated by lower viscosity fluid at the walls, a consequence of wall shear \citep{nolan2016}. As the fluid approaches the cylinder face, being forced to accelerate into the gaps, shear from the velocity gradient lowers its viscosity. From then on, travelling downstream, we can notice two low-viscosity regions. The low-viscosity region near the walls is again due to wall shear, while the second region is due to the velocity variations felt between the high-viscosity, high-velocity fluid leaving the gaps and the high-viscosity, slow-moving downstream wake. Increasing $Wi$ leads to significant changes upstream of the cylinder as the upstream wake grows, taking the form of a slow-moving, high-viscosity slug. Low-viscosity regions between the upstream wake and the faster-moving fluid travelling near the walls into the gaps can be observed. For both channel designs, as the lateral flow asymmetry onsets ($200<Wi<300$ for WTI and $300<Wi<400$ for NTI) the low-viscosity region at the gap opposing the upstream pressure tap (positive-$y$) grows in intensity due to the more acute velocity gradients as the fluid abandons the preferential gap. 

The employed viscosity models (Carreau--Yasuda) only consider the fluid's response to shear and do not contemplate elasticity. The lateral flow asymmetry instability (for the case of small blockage ratios, according to Hopkins et al.\citep{hopkins2022}) is thought to be due to a fluctuation in the highly stressed downstream wake and fed by the shear-thinning behaviour of the fluid. Instead of analysing the extensional viscosity of the fluids in the strain-hardened downstream wake, it was more relevant, in our view, to observe the shear-thinning effects, as it should be the characteristic responsible for actually maintaining the asymmetry. Moreover, both fluids are expected to behave similarly in the same extension-dominated flow regions, as both are elastic fluids subject to the same Weissenberg numbers at low Reynolds numbers. This is true within our observation area as both fluids similarly develop extended up- and downstream wakes.

\begin{figure}[htp]
    \centering
    \includegraphics[width=.8\textwidth]{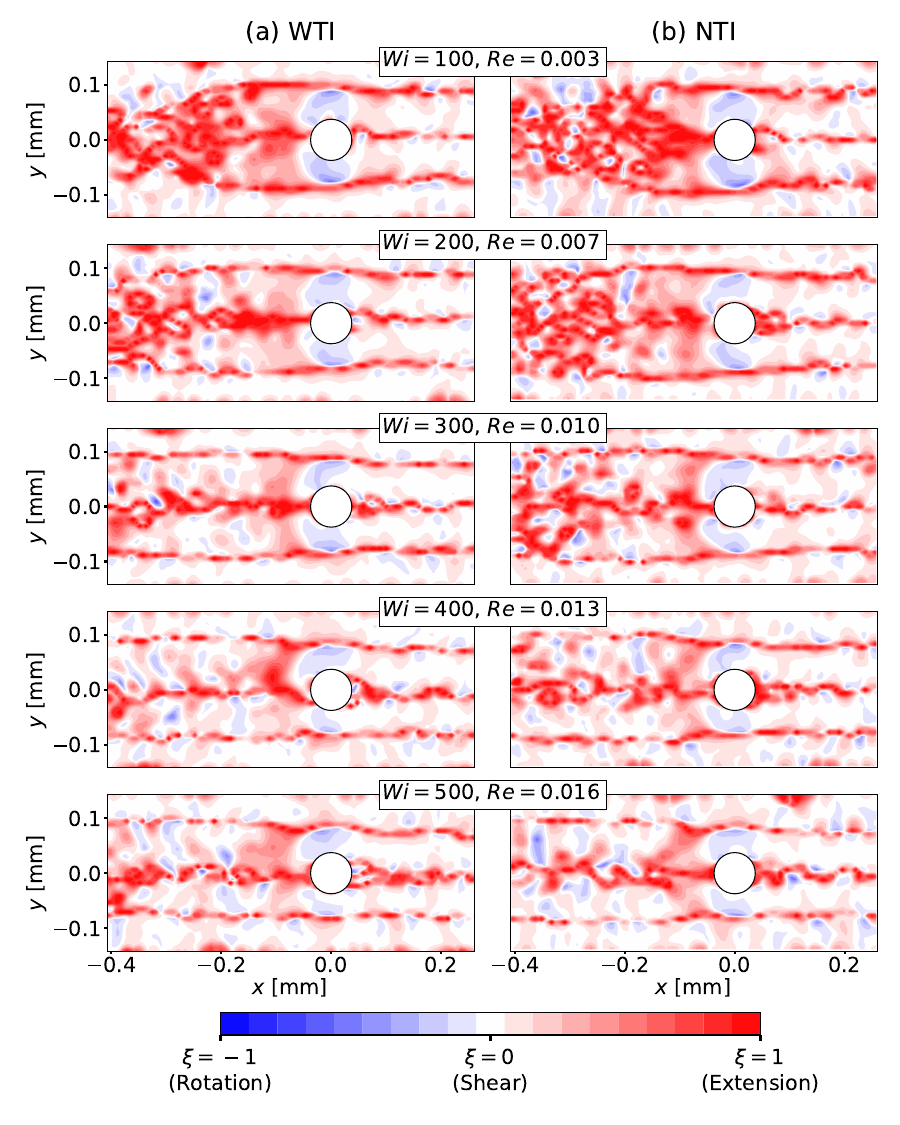}
    \caption{Time-averaged flow-type parameter flow fields, $\xi$, computed for (a) WTI and (b) NTI channels for PEO8 (flow is left to right).}
    \label{fig:flowtype PEO8}
\end{figure}

The flow-type parameter flow fields obtained with PEO8 are shown in Figure \ref{fig:flowtype PEO8}. Upstream of the cylinder, PEO8 shows larger, and more complex regions of extensional flow than PEO4. The three streams of extensional flow, which are related to the gaps and the stagnation points, are still present and become longer with $Wi$ upstream of the cylinder as the large extensional flow region is replaced with shear flow. This progressive replacement of the extensional flow with the shear flow and the growth of the extensional streams seems to occur slower than with PEO4. The rotational flow areas located in the gaps, near the cylinder face also remain for the shear-thinning fluid. Once lateral asymmetry onsets, there are noticeable differences between PEO8's WTI and NTI flows near the forward stagnation point. For the WTI flow, a high extensional flow region appears strongly at the positive-$y$ portion of the channel, where the fluid diverges mainly into the positive-$y$ gap, while on the NTI channel, it does not seem to be as intense (particularly noticeable for $Wi=400$).

\begin{figure}[htp]
    \centering
    \includegraphics[width=.75\textwidth]{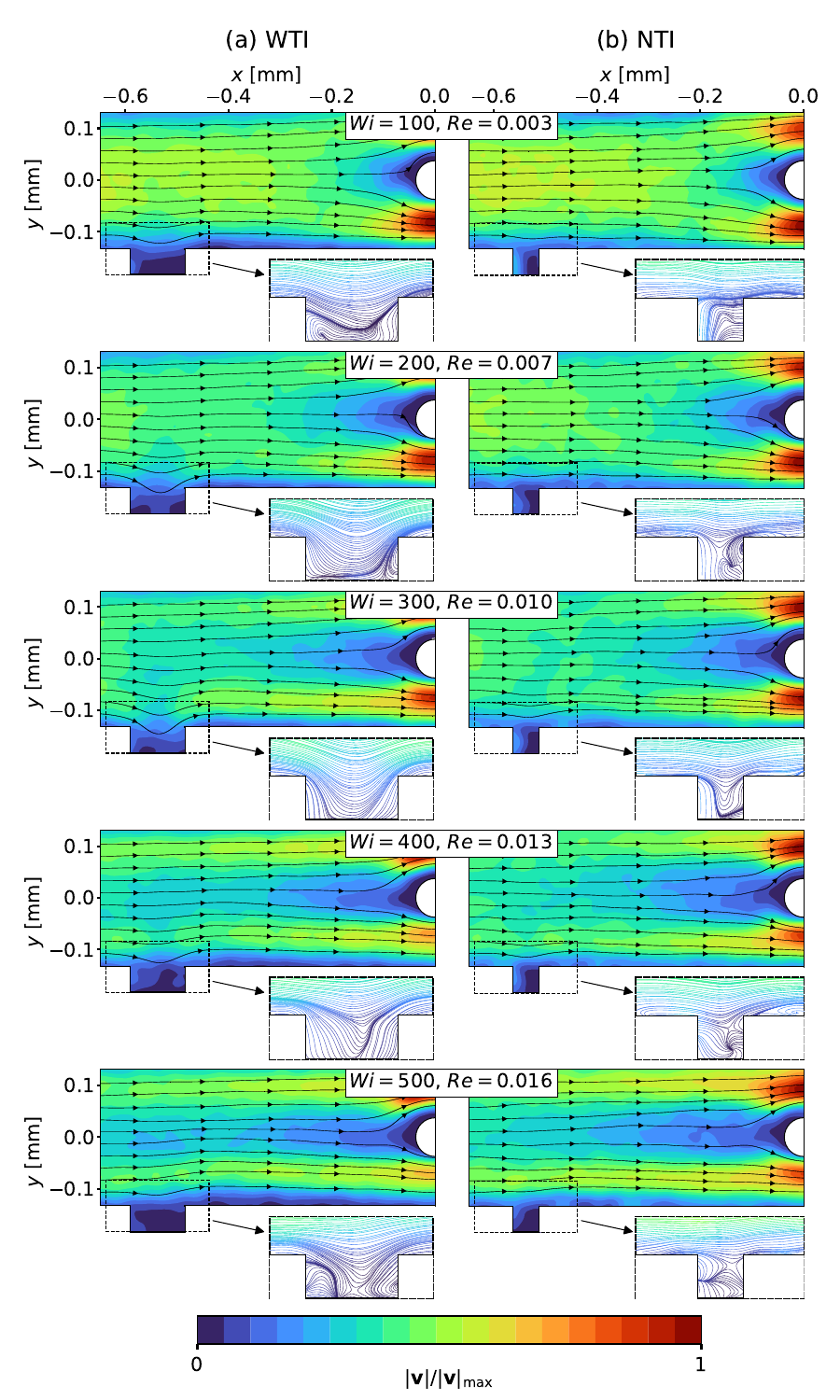}
    \caption{Time-averaged velocity magnitude flow fields, $\lvert \mathbf{v} \rvert$, and superimposed streamlines obtained with PEO8 for (a) WTI and (b) NTI channels, past the upstream pressure tap intake (flow is left to right).}
    \label{fig:Taps insets PEO8}
\end{figure}

The velocity magnitude fields at the intakes were plotted and are shown in Figure \ref{fig:Taps insets PEO8}. Again, we see the influence of the pressure taps on the flow with the larger intake allowing for a deeper dividing streamline and affecting the surrounding flow more significantly, as can be gathered from the streamline curvature near it. Overall, the behaviour at the tap intakes is similar to the one obtained with the weakly shear-thinning PEO4. Despite the lateral flow asymmetry, the NTI does not seem to influence the flow significantly. The flow-type parameter flow fields at the intake (shown as supplementary material) also show similar results to PEO4, with strong extensional and rotational flow regions near the upstream intake, connected to the deeper penetration of the fluid into the tap. The slower substitution of the large extensional flow region with shear flow, and the growth of the extensional flow streams upstream, is also clear, as displayed in Figure \ref{fig:flowtype PEO8}.

\begin{figure}[htp]
\centering
\begin{minipage}[t]{.40\textwidth}
    \centering 
    PEO8
    \\
    \footnotesize{($Wi = 300$, $Re = 0.010$)}
    \\
    \footnotesize{(a) WTI}
    \includegraphics[ width=\linewidth]{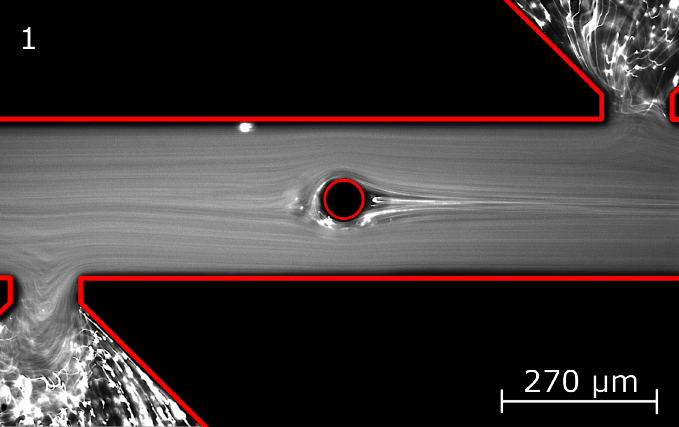}
    \includegraphics[ width=\linewidth]{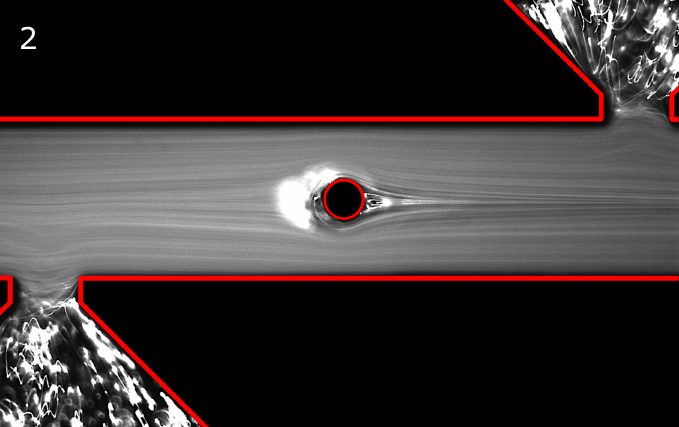}
    \footnotesize{(c) NTI}
    \includegraphics[ width=\linewidth]{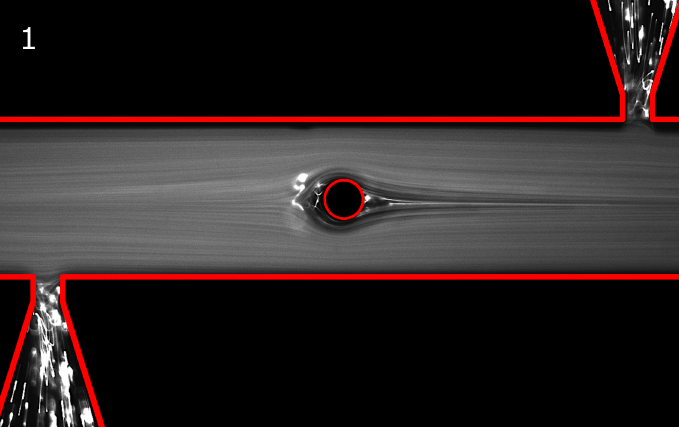}
    \includegraphics[ width=\linewidth]{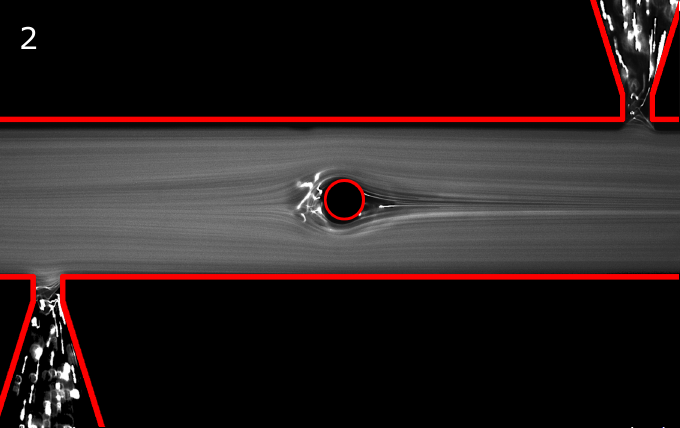}
\end{minipage}
\begin{minipage}[t]{.40\textwidth}
  \centering
    Water
    \\
    \footnotesize{($Wi = 0$, $Re = 0.010$)}
    \\
    \footnotesize{(b) WTI}
    \includegraphics[ width=\linewidth]{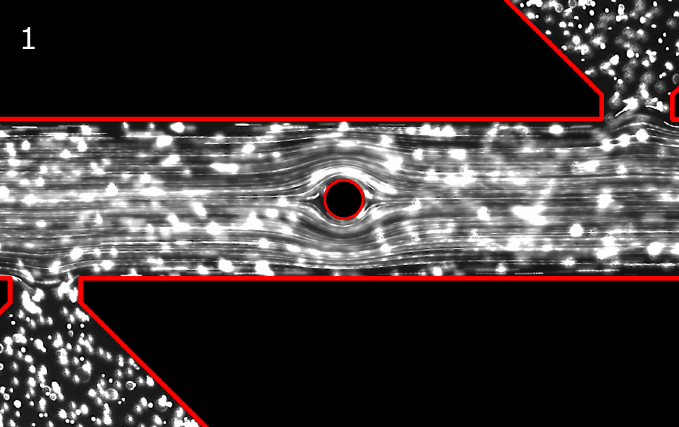}
    \includegraphics[ width=\linewidth]{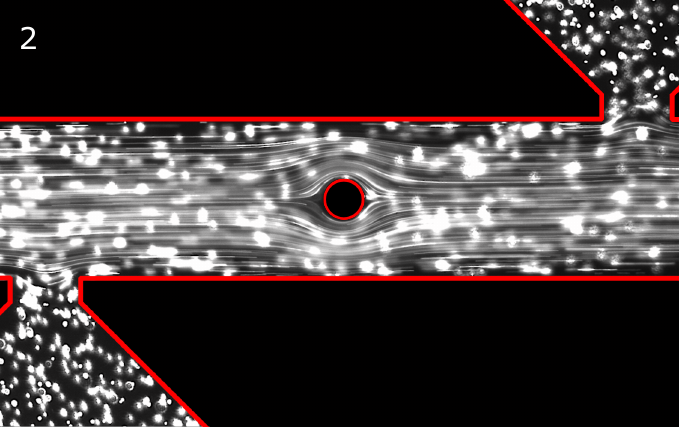}
    \footnotesize{(d) NTI}
    \includegraphics[ width=\linewidth]{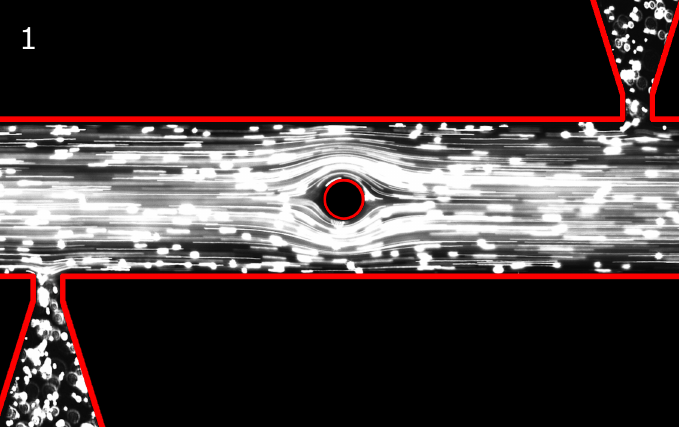}
    \includegraphics[ width=\linewidth]{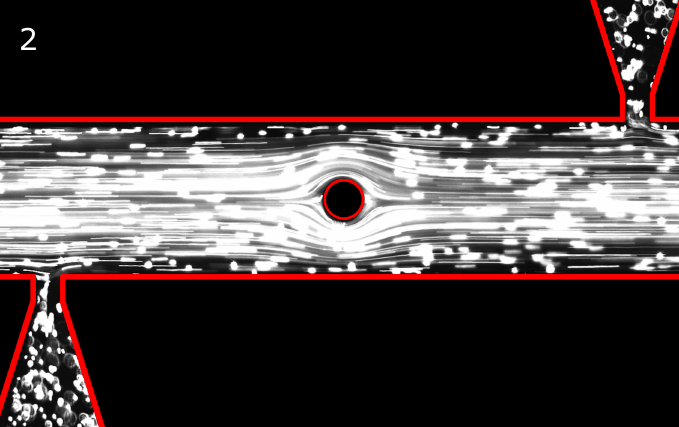}
\end{minipage}
\caption{Pairs of consecutive streak images (1 and 2), obtained with (left) PEO8 and with (right) Water, on ((a) and (b)) WTI and ((c) and (d)) NTI channels (flow is left to right).}
\label{fig:Streak PEO8}
\end{figure}

Once more, in order to evaluate the instantaneous flow shapes, flow visualisations were conducted with PEO8 at $Wi=300$ and with water at the corresponding Reynolds number. The resulting streak images can be seen in Figure \ref{fig:Streak PEO8}. Firstly considering the WTI flows (Figure \mbox{\ref{fig:Streak PEO8} (a)}), the two PEO8 images show that the location of the dividing streamline within the tap intake is related to the degree of lateral flow asymmetry around the cylinder, as was previously discussed for PEO4. Again, the streamlines begin to bend slightly far upstream of the cylinder near the tap intake, now with the absence of the wineglass ``pinch'' as it approaches the stagnation point. The flow is steady for $Wi=100$ and slight modifications of the intake vortex appear when increasing $Wi$ to $200$. Only when $Wi$ is increased to $300$ do we find such time-dependent behaviour as shown in Figure \ref{fig:Streak PEO8}(a). Increasing $Wi$ leads to transient motion of the forward stagnation point, detaching and collapsing back into the cylinder face, similar to the behaviour reported by Kenney et al.\cite{kenney2013}. With the NTI channel, no streamline bending as significant as with the WTI (\mbox{Figure \ref{fig:Streak PEO8} (a-1)}) is found. The flow remains steady and symmetric for the least elastic flow regimes. Increasing $Wi$ from 200 to 300 seems only to cause a slight detachment of the forward stagnation point. For $Wi = 400$, the transient vertical motion of the dividing streamline within the tap intake is noted, and so is time-dependent lateral flow asymmetry around the cylinder (agreeing with the \textmu PIV results). Moreover, the forward stagnation point again shows a transient horizontal shift. Increasing $Wi$ to 500 enhances the previously mentioned behaviour, with the forward stagnation point shifting even further upstream and showing transient horizontal and vertical motion, being the latter still in accordance with the penetration of the diving streamline. The Newtonian flow remained perfectly symmetric and steady for all tested flow regimes, for both channel designs. Additional visualisations were conducted with PEO4, at the same Reynolds number, $Re = 0.01$, however, the flows remained steady and symmetric with both channel designs (data not shown).

It is indeed curious that lateral flow asymmetry is found for both channel designs with the shear-thinning fluid. While the velocity magnitude flow fields do not blatantly reveal many differences between the two channel flows, except for the critical Weissenberg number for the onset of flow asymmetry, other analyses reveal differences between the two behaviours. The streamwise velocity profiles show that with the NTI lateral flow asymmetry is found only near the cylinder and downstream, while with the WTI, it can already be found immediately after the upstream tap intake. Moreover, the flow-type parameter flow fields show a few differences in the nature of the flow near the forward stagnation point, where the fluid diverges into the gaps, with the WTI yielding a more intense extensional flow in the positive-$y$ region of the channel. Lastly, less pronounced streamline curvature near the upstream NTI, can be seen either in the \textmu PIV results or the streak images. All these observations hint at a fundamental difference between the two flow asymmetries. The lateral flow asymmetry obtained with PEO8 on the NTI channel, could be due to the fluid's larger elasticity ($El_\mathrm{PEO8}=30323$ and $El_\mathrm{PEO4}=620$), which makes it more susceptible to flow perturbations --- Cochrane et al.\citep{cochrane1981} concluded that ``symmetry is elusive for elastic liquids whenever there is a mechanism in the flow for promoting an asymmetry of any degree'' (p.170)). However, the evidence found regarding the differences between the two asymmetric flow behaviours leads us to believe that the taps and elasticity alone should not be responsible for the phenomenon on the NTI channel, contrarily to PEO4's behaviour with the WTI. It could be thought that, similarly to other reported lateral flow asymmetries\cite{nolan2016,ribeiro2014,zhao2016,haward2019,haward2020,varchanis2020,hopkins2022}, it is indeed due to a viscoelastic instability, onset by elasticity and fed by the fluid´s strong shear-thinning. Nevertheless, the fact that the fluid chooses the positive-$y$ gap as a preferential path around the cylinder (which opposes previous reports of the fluid choosing the path seemingly at random\cite{haward2020,varchanis2020}) and that the movement of the forward stagnation point seems to be connected to the depth of penetration of the dividing streamline within the intake, point towards the taps' influence. Haward et al.\cite{haward2019} proposed the possibility of the viscoelastic instability's onset mechanism being a random fluctuation in the highly stressed downstream wake that provokes a momentary flow rate imbalance between the two cylinder/wall gaps, which, due to the consequent shear rate imbalances and the fluid's shear-thinning, leads to the maintenance of a path of least resistance around the cylinder. In our case, it is possible that, instead of the fluctuation in the wake, the NTI is unsettling the flow, serving as the onset mechanism, and elasticity and shear-thinning are feeding the asymmetry.

The set of flow transitions encountered in this study agrees relatively well with the work of Hopkins et al.\citep{hopkins2022}, except that the critical Weissenberg numbers for the onset of our lateral flow asymmetry and time-dependent flow are larger (Hopkins et al.\citep{hopkins2022} reports critical $Wi\approx10$ for steady asymmetric flow and $Wi\approx40$ for time-dependent flow, for a similar blockage ratio). Moreover, our maximum degree of asymmetry achieved ($I\approx 0.19$) is lesser than the values obtained by Varchanis et al.\cite{varchanis2020} ($I\approx0.7$), who studied the same fluid with a lower blockage ratio ($\beta=0.1$). Both studies found that increasing channel blockage (within an upper boundary, in the case of low blockage channels\citep{hopkins2022}) should facilitate the onset of asymmetric flow. One possible explanation for these discrepancies lies in the fact that our microchannels have much lower aspect ratios ($\alpha=0.37\ll5$) and, according to the work of Ribeiro et al.\cite{ribeiro2014}, the increased confinement of the channel should delay the lateral flow asymmetry's onset. Moreover, the fact that the magnitude of our asymmetries was not as large could simply be due to our flow not achieving sufficiently large Weissenberg numbers, as we did not observe the subsequent partial resymmetrisation of the flow\cite{haward2019,haward2020,varchanis2020}. Alternatively, as we are dealing with transient flow and our \textmu PIV results are time-averaged, more intense lateral flow asymmetries could indeed be occurring.
\pagebreak

\section{Conclusions}
\label{sec:conclusions}

In this work, the flow of two polymeric solutions (one elastic weakly shear-thinning --- PEO4, and one elastic shear-thinning --- PEO8), around a confined microfluidic cylinder was studied on microchannels containing pressure taps which differed on the intake width (Wide-Tap-Intake --- WTI, and Narrow-Tap-Intake --- NTI). Previous works on the flow of viscoelastic fluids past confined cylinders have encountered laterally asymmetric flow, with more fluid flowing through either side of the cylinder\cite{nolan2016,ribeiro2014,zhao2016,haward2019,haward2020,varchanis2020,hopkins2022}. The most recent hypothesis claims it is due to a viscoelastic instability, which requires both fluid elasticity and strong shear-thinning for its onset\citep{varchanis2020}. In this study, laterally asymmetric flow was encountered not only with the elastic shear-thinning PEO8 but also with the elastic weakly-shear-thinning PEO4, which goes against these previous reports. For the latter, it was only found on the WTI channel, which leads us to believe that it should not be due to the known viscoelastic instability but, instead, to a flow perturbation provoked by the presence of the wider pressure tap.

Contrarily, with the shear-thinning PEO8, lateral flow asymmetry was found on both channels. However, there was evidence of flow asymmetry far upstream of the cylinder, only near the WTI. Therefore, on the NTI channel, the lateral flow asymmetry found with PEO8 could indeed be due to the previously reported viscoelastic instability.

Time-dependent flow was also observed at high Weissenberg numbers, evident from the fluctuations of the positioning of the forward stagnation point, seemingly connected with the vertical oscillations of the dividing streamline inside the upstream pressure tap intakes. As unsteady flow has been previously reported independently in viscoelastic flows past confined cylinders\citep{kenney2013,zhao2016,hopkins2022,haward2019,varchanis2020} and past cavities\citep{kim2000}, we cannot assertively conclude on the nature of our time-dependence.

Nevertheless, the connection between the degree of lateral flow asymmetry around the cylinder and the penetration of the dividing streamline into the tap intake may further imply that the pressure tap may be causing the flow to ``trip'', perhaps even be provoking the flow asymmetry around the cylinder itself. Fluid elasticity\citep{haward2018,cochrane1981} and tap intake width seem to be key parameters for the magnitude of the flow perturbation.

Pressure tap design and its influence on microfluidic elastic flows is still a relatively untouched subject, and, to the best of our knowledge, this work stands as the first study dedicated to the topic, despite merely scratching its surface. Including only the upstream pressure tap or removing them altogether from the microchannel, could shed some light on the tap's influence on the flow and the nature of the encountered lateral flow asymmetry. Moreover, conducting pressure measurements could grant additional information, not only on the transitions to steady asymmetric flow and the characteristics of the subsequent time-dependency, but also on the suitability of the pressure tap designs and the associated hole-pressure error.

The small length-scale associated with microfluidics enhances the flow's viscoelastic characteristics, making pressure measurements a task requiring careful consideration to avoid flow interference. A compromise between optimal pressure measurements and flow disturbance is achievable but requires further investigation and understanding of the viscoelastic flow dynamics.

\pagebreak

\section*{Supplementary Material}

As supporting material, we present the flow-type parameter plots obtained with PEO4 and PEO8 on both channel designs, which allow the visualisation of the flow near the upstream tap intakes (briefly discussed in the Results). Additionally, we have included local viscosity plots obtained with PEO4 and a brief discussion on the employed normalisation formulation.

\section*{Acknowledgements}

This research was funded by FEDER (COMPETE 2020) and FCT/MCTES (PIDDAC) under grants Nos. PTDC/EME-EME/30764/2017, PTDC/EME-APL/3805/2021, 2022.08598.PTDC, LA/P/0045/2020 (ALiCE), UIDB/00532/2020 and UIDP/00532/2020 (CEFT). T.R. acknowledges FCT for funding support under Scholarship No. 2021.06532.BD.





\bibliographystyle{unsrtnat}
\bibliography{bibliography.bib}


\newpage
\pagestyle{plain}

\renewcommand{\thepage}{S\arabic{page}}
\renewcommand{\thesubsection}{S\arabic{subsection}}
\renewcommand{\thetable}{S\arabic{table}}
\renewcommand{\thefigure}{S\arabic{figure}}
\renewcommand{\theequation}{S\arabic{equation}}

\renewcommand\bibsection{\subsection*{\refname}}
\renewcommand{\bibnumfmt}[1]{[S#1]}
\renewcommand{\citenumfont}[1]{S#1}

\setcounter{page}{1}
\setcounter{subsection}{0}
\setcounter{table}{0}
\setcounter{figure}{0}
\setcounter{equation}{0}




\end{document}